\documentclass[prl,twocolumn,english,showpacs]{revtex4-1}
\usepackage[utf8]{inputenc}
\usepackage[english]{babel}
\usepackage{amsmath}
\usepackage{amsfonts}
\usepackage{amssymb}
\usepackage{lipsum}
\usepackage{graphicx}
\newcommand{\ket}[1]{\left|#1\right\rangle}

\newcommand{\element}[1]{${}^{#1}$}
\newcommand{\spectralLvl}[3]{${}^{#1}\text{#2}{}_{#3}$}

\begin{document}
	
	\title{Deterministic spin-photon entanglement from a trapped ion in a fiber Fabry-Perot cavity}
	
	\author{Pascal Kobel, Moritz Breyer and Michael K\"ohl}
	
	\affiliation{Physikalisches Institut, University of Bonn, Wegelerstra{\ss}e 8, 53115 Bonn, Germany}
	
	\begin{abstract}
		The development of efficient network nodes is a key element for the realisation of quantum networks which promise great capabilities as distributed quantum computing or provable secure communication.  We report the realisation of  a quantum network node using a trapped ion inside a fiber-based Fabry-Perot  cavity. We show the generation of deterministic entanglement at a high fidelity of $ 91.2(2) $\,\% between a trapped Yb--ion and a photon emitted into the resonator mode. We achieve a success probability for generation and detection of entanglement for a single shot of $ 2.5 \cdot 10^{-3}$ resulting in 62\,Hz entanglement rate. 
	\end{abstract}
	
	\maketitle

\section*{Introduction}

The communication between different quantum nodes is central to many branches of quantum technology, such as quantum computation \cite{QuantumInternet}, quantum metrology \cite{RevModPhys.89.035002} and secure communication \cite{NonCloningTheorem,QuantumKeyDistrBB84}. Quantum nodes consist of an interface between a stationary qubit and a traveling qubit, which usually is a photon in order to bridge large distances. Therefore, the development of light-matter interfaces operating at the quantum limit and providing control of the quantum states of both the traveling qubits and the quantum network node itself is of great importance. The large range of possible applications leads to a broad spectrum of requirements for possible quantum network nodes. Firstly,  nodes have to exhibit long coherence times, scaling with the characteristic distances of the targeted network. Secondly, the stationary qubits at the network nodes must be efficiently coupled to traveling qubits which carry quantum information across the network  application. Thirdly, the stationary qubits should be accessible to a high-fidelity readout. There have been several approaches towards the experimental realization of quantum network nodes, for example, using trapped ions \cite{Blinov2004,2012_fastest_spin_photon,Bock2018,Stute2013}, neutral atoms \cite{Wilk488,PhysRevLett.96.030404}, NV-centers \cite{Togan2010} or SV-centers \cite{PhysRevLett.123.183602} in diamond, and semiconductor quantum dots \cite{DeGreve2012,Gao2012}. For both the photon collection and the readout of the network node, experimental approaches have ranged from  collecting light with high numerical aperture objectives \cite{PhysRevLett.124.110501,Togan2010,PhysRevLett.103.213601,Gerber_2009} to embedding emitters in optical cavities in order to enhance the light-matter coupling strength \cite{Wilk488,Krutyanskiy2019,2012_fastest_spin_photon,PhysRevLett.110.043003}. 
In order to distribute quantum information in networks or to deliver it to frequency-conversion units, it is desirable to couple travelling photonic qubits into optical fibers.  A particularly elegant approach towards this goal are fiber-based Fabry-Perot resonators \cite{Hunger_2010}. This device composes of a pair of optical fibers with micromachined endfacets with a highly-reflective optical coating. They combine the best of two worlds: on the one hand, they are optical microresonators with small mode volume and on the other hand the cavity mode is directly fiber-coupled. Cavities with small mode volumes offer large coupling between light and matter and therefore can enhance the efficiency of the light-matter interface. Moreover, they have the conceptual advantages of overcoming difficulties of orthogonality of polarizations in light collection with high numerical aperture objectives due to their small solid angle and  increasing the bandwidth of the emitted photons. The integration of fiber Fabry-Perot cavities into quantum network nodes is considered to be a promising route for quantum communication \cite{Northup2014}.  
However, even though  neutral atoms \cite{PhysRevLett.121.173603,Colombe2007}, trapped ions \cite{PhysRevLett.110.043003,MSteinerMeyer,PhysRevA.96.023824,doi:10.1063/1.4838696,Takahashi2020}, NV centers \cite{PhysRevLett.110.243602} and semiconductor quantum dots \cite{doi:10.1063/1.3245311} have been coupled to fiber Fabry-Perot resonators, the entanglement between light and matter in such an experimental setting has not yet been demonstrated.

In this work, we  realize a quantum network node of a trapped \element{171}Yb\element{+} ion coupled to a fiber Fabry-Perot cavity and demonstrate the generation and verification of a  maximally entangled atom-photon state with a fidelity of $ F= (91.2\pm0.2)\text{ \%} $. The entanglement is realised between the spin state of the trapped ion and the polarisation degree of freedom of the emitted photon. We have measured a single-shot success probability of $ 2.5\cdot10^{-3} $ for generation and detection of entangled states resulting in a rate of $62 \text{\,s}^{-1} $.


\section*{Results}

Our quantum network node consists of a single trapped $^{171}$Yb$^+$ ion in a radiofrequency (Paul) needle trap embedded into a fiber Fabry-Perot cavity, conceptually similar to our previous work \cite{PhysRevLett.110.043003,MSteinerMeyer,CavityBackAction}, see Figure 1a. The ion is confined with trap frequencies of $\omega_{x,y}=2\pi\cdot 1.5 \text{\,MHz}$ and $\omega_z=2\pi\cdot2.5 \text{\,MHz}$ and Doppler-cooled by a near-resonant laser at 370\,nm wavelength. The  fiber Fabry-Perot cavity has a length of  261\,$ \mu $m and a finesse of $ F=4700\pm 700 $ at 370\,nm and serves as a light-matter interface on the principal $^2S_{1/2}\,\leftrightarrow \,^2P_{1/2}$ dipole transition of the ion. The cavity linewidth is $ \kappa= 2\pi\cdot\left(58 \pm 9\right)$\,MHz and therefore photons emitted by the ion and leaving the cavity exhibit a duration of $ 9.9(6)$\,ns (FWHM).

For the generation of entanglement, we initialize the ion  in the hyperfine ground state $\ket{0}=\ket{^2S_{1/2},F=0,m_F=0}$  within 10\,$\mu$s with more than $99\% $ fidelity using optical pumping with continuous-wave laser light. Subsequently, we excite the ion to the electronically excited state $\ket{e}=\ket{^2P_{1/2},F'=1,m_F=0}$ using  a laser pulse derived from a frequency-doubled mode-locked Ti:sapphire laser of duration $ t_{\text{pulse}}=(74\pm 2 )$\,ps, see Figure 1b. The pulse duration is much shorter than the lifetime of the  $^2P_{1/2}$ state of  $ \tau=1/\Gamma^{'}=7.39(15) \text{ ns} $. The  laser pulse intensity is calibrated by driving Rabi oscillations on the $S\leftrightarrow P$ transition and we observe a fidelity of the preparation of the $\ket{e}$ state in excess of $97\%$. Following the excitation, the ion decays in a superposition of decay channels into the $ \ket{g^+}=\ket{^2S_{1/2},F=1,m_F=-1} $ and $ \ket{g^-}=\ket{^2S_{1/2},F=1,m_F=1} $ states by emission of a single photon. We apply an external magnetic field of 603.6(7)\,mG  along the cavity axis in order to suppress  emission into the cavity mode with a change of magnetic quantum number of $ \Delta m_F=0 $. Hence, only circular polarized photons are emitted into the cavity mode. The success probability of an emission into the cavity mode is  $(10.1\pm1.9)\%$, see Methods. The external magnetic field leads to a level shift between the $\ket{g^\pm}$ states, however, the frequency difference between the two polarization modes $ \sigma^\pm $ is much smaller than the atomic linewidth  which preserves the capability of the ion--cavity system to decay in a superposition of channels.  Hence for a photon emitted into the cavity mode the ideal maximal entangled atom-photon state reads $\ket{\Psi_{\text{atom-photon}}}=\frac{1}{\sqrt{2}}\left(\ket{\sigma^+}\ket{g^+}-\ket{\sigma^-}\ket{g^-}\right)$.

The collected photons leave the cavity into a single-mode optical fiber.  After $ \sim 1.5\text{\,m} $  of fiber we detect the quantum state of the photons by a projective measurement on a polarising beam splitter (PBS) and two single photon counters (SPCs), one on each exit path of the PBS. The photonic readout basis is defined by a set of waveplates which rotate the polarisation qubit into an arbitrary basis (see Methods and Figure \ref{wavePlateCavitySimulation}). After the polarization rotation, we end up with the transformed state
 \begin{equation}
	\ket{\Psi_{\text{atom-photon}}}=\frac{1}{\sqrt{2}}\left(\ket{V}\ket{g^+}-\ket{H}\ket{g^-}\right), 
 \end{equation} 
in which $\ket{H}$ and $\ket{V}$ denote the horizontal and vertical polarisations, respectively. 

\begin{figure*}
 	\centering
 	\includegraphics[width=0.9\textwidth]{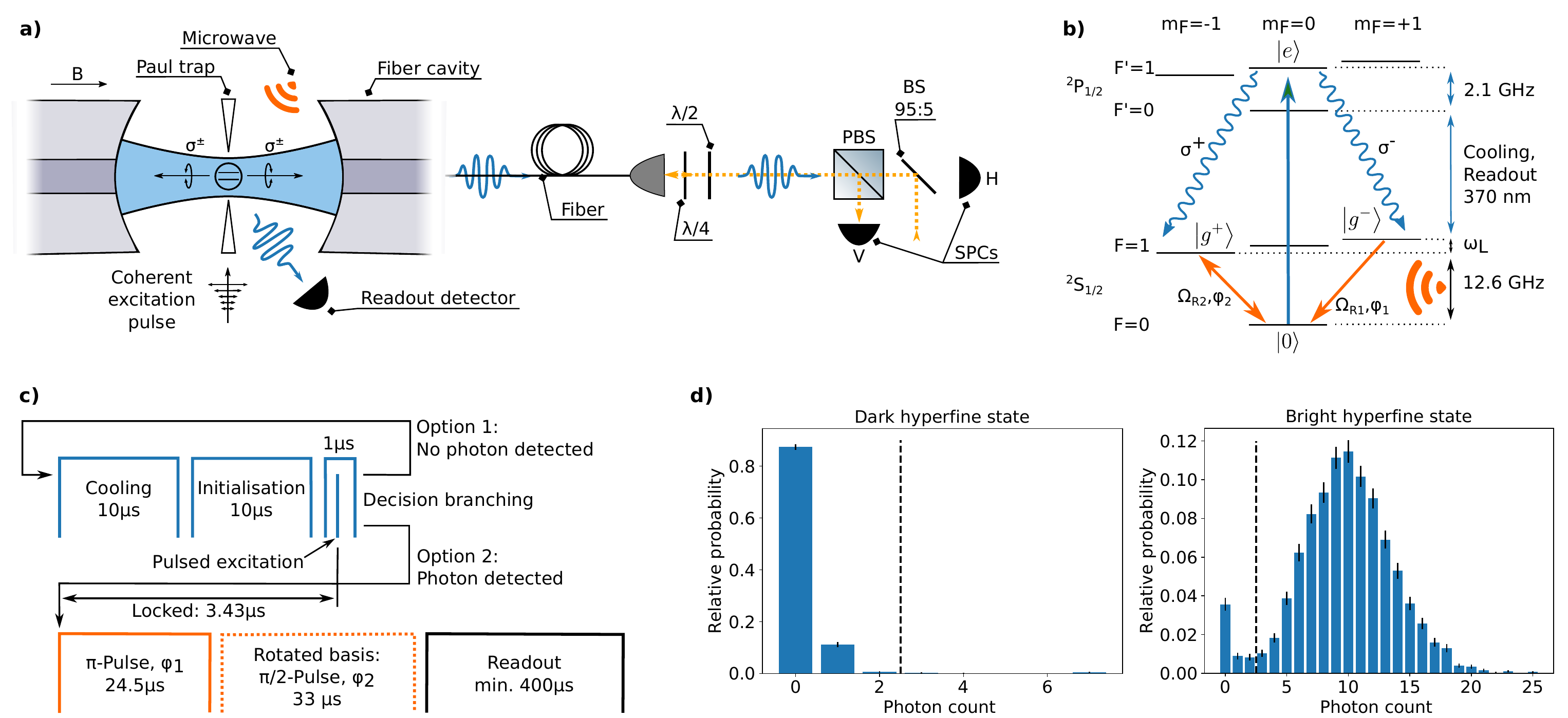}
 	\caption{ {\bf Experimental setup}  a) Experimental setup. We collect the single emitted photons along the quantisation axis with a fiber Fabry-Perot cavity and analyse them using a projective measurement of polarisation.
 		 b) Selected energy levels of \element{171}Yb\element{+} including the relevant optical transitions. After pulsed excitation from the $\ket{0}$ to the $ \ket{e} $ state, the ion decays in a superposition of decay channels emitting a $ \sigma^{\pm} $ polarised photon. Subsequent manipulation of the atomic qubit is done via microwave pulses. c) Experimental sequence for creating and verifying atom-photon entanglement including decision branching, see main text for explanation. d) Readout of the hyperfine qubit using discrimination between dark and bright hyperfine states. We achieve a readout fidelity of $96.9(3)$\,\% using the dashed line as bright/dark threshold. }
 	\label{lvlschemeAndQubitManipulation}
 \end{figure*}

After excitation of the ion and subsequent emission of the photon, we wait for a count on the single-photon counters. If no photon is detected within 1\,$\mu$s after the excitation, the experimental sequence continues with new initialization and excitation of the ion qubit (see Figure \ref{lvlschemeAndQubitManipulation}c).  In contrast, if a photon has been detected, we read out the atomic state in order to conduct a correlation analysis between photon polarization and atomic spin state. The first stage of detection is the mapping from the Zeeman qubit states $\ket{g^\pm}$ to the hyperfine states $\ket{g^+}$ and $\ket{0}$. To this end, we employ a microwave $ \pi $-pulse from the  $\ket{g^{-}} $ to the $\ket{0}$ state at a frequency near 12.6\,GHz.  Considering the ion as a two-level system, we refer to this atomic basis as the $\sigma_z$--basis. This is then followed by hyperfine-state selective fluorescence detection. We are able to discriminate between the hyperfine qubit states within a few 100\,$ \mu $s by using fluorescence detection on the $\ket{^2S_{1/2},  F=1} \leftrightarrow \ket{^2P_{1/2},  F'=0}$ transition. For detection of the photons emitted into  free space we use an objective with a numerical aperture of  $ \text{NA}=0.48 $.
The distribution of photon numbers for a dark and bright ion are shown in Figure 1d for 400\,$ \mu $s readout time. In total, we achieve a discrimination fidelity between a dark and a bright ion of 96.9(3)\% mainly limited by off resonant scattering of photons on the $ \ket{^2P_{1/2},  F'=1} $ level for a bright ion.  

In Figure 2a we show the results of correlations between photons in the original polarization states $\sigma^\pm$ and the atomic states $\ket{g^+	}$ and $\ket{0}$. We observe correlations with $(90.7\pm3.9)$\,\% contrast which is mainly limited by state discrimination $ 6.2(3)\text{\,\%} $. 

 In contrast to a statistical mixture of states, the entangled state of two qubits exhibits a definite phase relation between them, and quantum  correlations between atom and photon states are visible in any two-qubit basis $ \sigma_i\otimes\sigma_i $. In order to verify the entanglement in different bases orthogonal to $ \sigma_z $, we rotate both qubits into the equatorial plane of the Bloch sphere. For the photonic qubit we adjust the waveplate angles as described in Methods. For the atomic basis we apply a $ \pi/2 $--pulse with a phase difference of $ \Delta \phi=\phi_1-\phi_2  $ with respect to the first $ \pi $--pulse. The relative phase difference $ \Delta \phi$ of the pulses sets the exact readout basis of the atomic qubit, which is especially important to select a specific basis such as $\sigma_x \otimes \sigma_x$ and $\sigma_y \otimes \sigma_y$. For these rotated bases also the Larmor precession of superposition spin states comes into play. 
Both pulses originate from an arbitrary waveform generator and are mixed to a carrier signal red detuned by $ \sim 8  \text{\,MHz} $ from the center of the two microwave transition frequencies. Mixing to the same carrier preserves the relative phase $ \Delta \phi $ between the pulses. Since the experimental sequence is not synchronised to the microwave carrier phase, the first $ \pi $-pulse starts with a random phase with respect to the Larmor precession of the atomic qubit. The following $ \pi/2 $ pulse acting on the $ \ket{0} $/$ \ket{g^+} $ qubit rotates around an axis with a fixed relative orientation to the phase of the $ \ket{0} $/$ \ket{g^+} $ superposition. The relative orientation of this rotation axis is determined by the phase difference $ \Delta \phi $. In total, both pulses rotate the ion qubit around a fixed axis regardless of the phase of the first pulse. We end up in total with a defined atomic basis for readout by considering a fixed timing of the pulses in the laboratory frame with respect to the Larmor precession. 
Using this technique we are able to precisely select any basis orthogonal to $ \sigma_z $ for the atomic qubit.

In Figure 2b, we show the oscillation of correlations between a photon readout in the basis $\sigma_x/\sigma_y$ and a varying atomic readout basis determined by $ \Delta \phi $, also referred to as parity oscillations. We observe correlations with $(81.3\pm15.8)\text{\,\%}$ contrast for $ \sigma_x \otimes \sigma_x $ and $ (87.0\pm2.6)\text{\,\%} $ for $ \sigma_y \otimes \sigma_y $. The contrast is limited by decoherence due to magnetic field noise $(5\pm3)\text{\,\%}$, timing jitter of the microwave pulses ($ \leq 5 $\,\%) and increased error in atomic state discrimination due to formation of coherent dark states in the readout transition (see Methods).

From the correlation measurements in $z$-basis and in the orthogonal bases we calculate the lower bound of the detection fidelity for the maximally entangled atom-photon state to be $ F\geq (89.6 \pm 1.5)$\,\% (see Methods for details). This value is a measure for the quality of the whole setup and not only of the generated entangled state, as it includes all experimental noise and errors like SPCs dark counts, polarization mixing effects, qubit manipulations, the atomic readout and the dephasing of the atomic qubit due to magnetic field noise. To estimate the quality of the generation of entanglement, we investigate the contribution of each error source to the fidelity starting with correction for the dark counts on the SPCs resulting in 1.1(4)\,\% increase of the measured entanglement fidelity. We benefit from the short temporal shape of our photon wave function, which enables a short acceptance window of 10\,ns for post-analysis. Further, we determine the full density matrix of the two-qubit state and use a maximum likelihood estimation \cite{PhysRevA.64.052312} to ensure the density matrix $ \rho $ to be physical, see Figure 3a. We use this matrix to determine the contribution of unwanted unitary rotations on the atom and photon bases to be 1.0(2)\,\%. This confirms our statement of having precise control over both qubit bases. Following \cite{Bock2018,Bussieres2014}, we compute an upper bound for the entanglement fidelity $ F_\text{max}=\frac{1}{2}\left(1+\sqrt{2P-1}\right) = \left(91.2\pm0.2\right)\text{\,\%}$ from the purity $ P=\text{Tr}\left(\rho^2\right)= 84.0(3)\text{\,\%}$ of the state.
This state fidelity is limited by noise contributions, for example qubit dephasing. Specifically, we have studied the influence of the external magnetic field  noise on the performance of our quantum network node. The external magnetic field has two main effects: (1) the magnetic field lifts the degeneracy between the $ \ket{g^-} $ and $ \ket{g^+} $ states which results in a precession of the atomic and photonic superposition states in the laboratory frame with Larmor-frequency $ \omega_L=2.8 \frac{\text{MHz}}{\text{Gauss}}\cdot (603.6\pm0.7) \text{ mG} = 1.690(2) \text{ MHz}$. For the successful detection of correlations, both atomic and photonic channels have to be stable with respect to this precession. For the photon, this requires a stable optical path length with fixed time between generation and detection. For the atom, the phases of the microwave pulses must have defined starting times relative to the flying qubit detection such that the pulses arrive always at the same phase of the Larmorprecession of the stationary qubit. We fix the starting time of microwave pulses with respect to the arrival time of the excitation pulse to less than 400\,ps by synchronising them to the cavity round-trip time of the Ti:sapphire laser. (2) Magnetic field fluctuations reduce the coherence time of the qubit of which the $\ket{g^\pm}$ states are magnetic-field sensitive. We have measured the coherence time of both the $\ket{g^\pm}$ and the $\ket{g^+}/\ket{0}$ qubits using a Ramsey sequence with variable hold time. For each hold time we obtain the remaining coherence fraction by the contrast of a full scan of the relative phase between the pulses. The results are shown in Figure 3b, and we observe a coherence time of $(496\pm42)\,\mu$s and $(1020\pm278)\,\mu$s for the $\ket{g^\pm}$ and the $\ket{g^+}/\ket{0}$ qubits, respectively. For the duration of the atomic basis manipulation we determine a reduction in contrast of $ (5\pm 3)$\,\%.


\begin{figure*}
	\centering
	\includegraphics[width=0.8\textwidth]{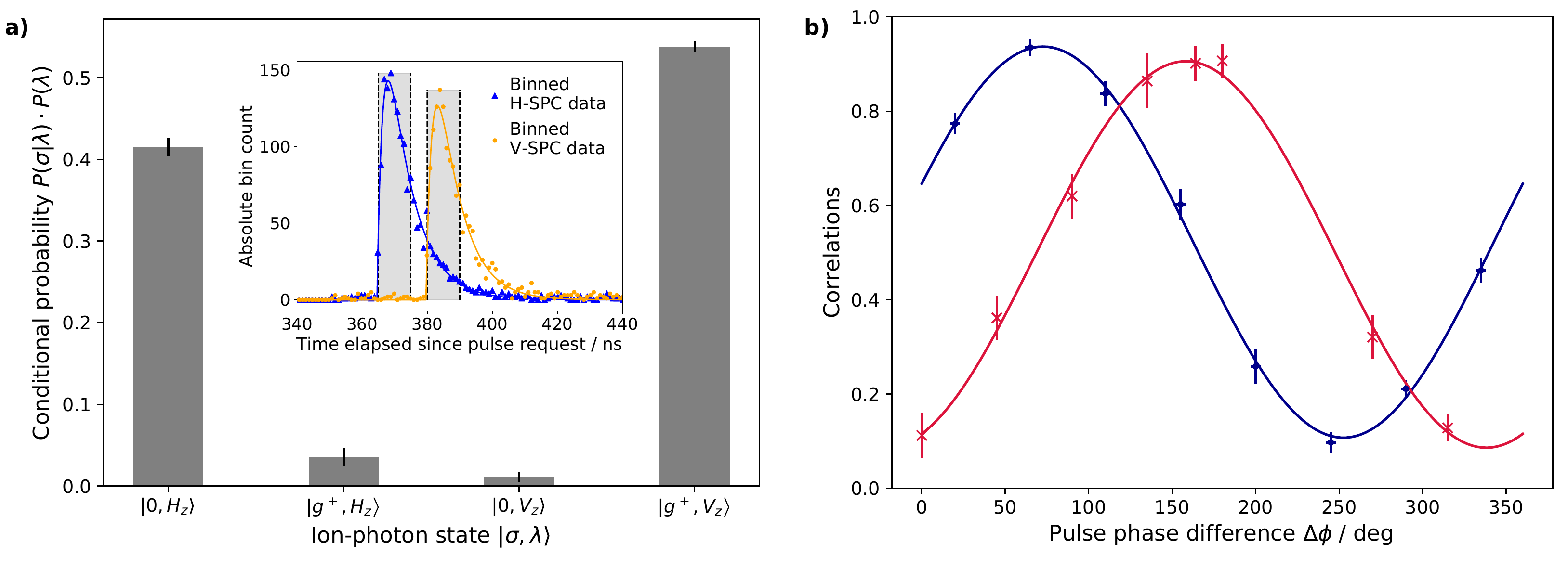}
	\caption{{\bf Verification of entanglement} a) Correlations for the $\sigma_z \otimes \sigma_z $ basis are shown with time-binned arrival times of the photons for this measurement as inset. We measure a FWHM of 9.9(7)\,ns for the photon duration. From the atomic decay constant the light matter coupling rate can be estimated (see Methods for details). The grey shaded areas represents the post selection boundaries (10ns width) for photons accepted for analysis (see main text for details).
		 b) The parity oscillations for a photon readout in $ \sigma_x $ and $ \sigma_y $ bases are shown.  The relative phase difference $ \Delta \phi=\phi_1-\phi_2 $ of the microwave pulses sets the readout basis of the atomic qubit in the equatorial plane of the Bloch sphere. Red crosses:  $P(0|H_x)\cdot P(H_x)+P(g^+|V_x)\cdot P(V_x)$ correlations. Blue dots: $P(0|H_y)\cdot P(H_y)+P(g^+|V_y)\cdot P(V_y)$ correlations. 
		}
	\label{Entanglement}
\end{figure*}

\begin{figure*}
	\centering
	\includegraphics[width=0.8\textwidth]{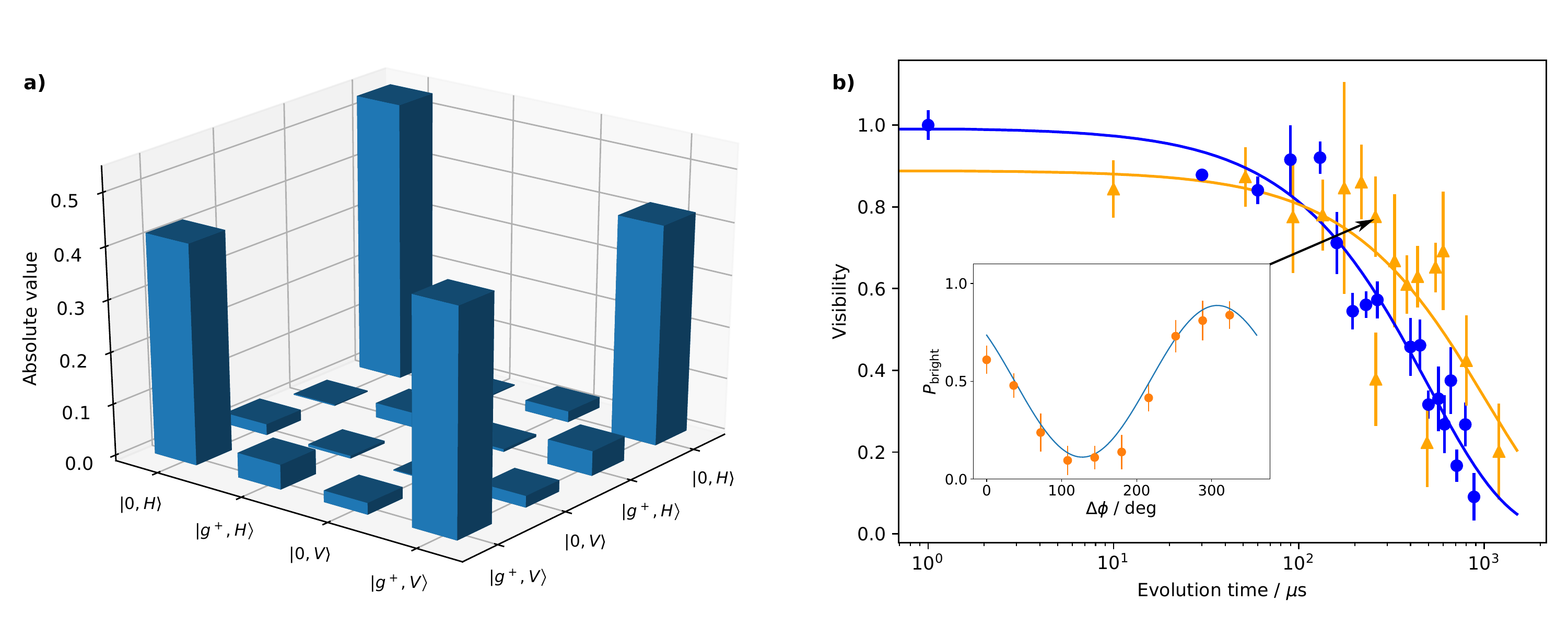}
	\caption{{\bf Full quantum state tomography and coherence time of the atomic qubit.} a) Absolute values of the density matrix $ \rho $ obtained from a full quantum state tomography followed by a maximum likelihood estimation as described in methods. b) The experimental data show the visibility of a Ramsey like sequence for different evolution times between the two $\pi/2$ pulses. The visibility for each evolution time is determined by scanning the relative phase between the pulses (inset). Blue circles: The Zeeman qubit $\ket{g^\pm}$ exhibits a coherence time of $(496\pm42)\,\mu$s. Yellow triangles: The hyperfine qubit $\ket{g^+}/\ket{0}$  exhibits a coherence time of $(1020\pm278)\,\mu$s. The times are extracted from fits (solid lines) with the function $ \exp\left(-\left(t/\tau\right)\right)$.}
	\label{Coherence}
\end{figure*}

\section*{Discussion}


The presented system is well suitable as a node in  quantum networks. The small size of the fiber Fabry-Perot cavity enables very good optical access to the trapped ion allowing for precise control and manipulation of the long-lived stationary qubit. We have achieved a deterministic entanglement at a high fidelity of  $ F_\text{max}=(91.2 \pm 0.2)$\,\%.
Effects reducing the fidelity are timing of the readout, error in state detection of the atomic qubit and dephasing of the atomic qubit due to magnetic field noise. The atomic transition linewidth of $ 2\pi\cdot\text{19.4\,MHz}$ allows for high generation rate of entanglement supported by fast extraction of photons out of the cavity of 1.3(2)\,ns which are intrinsically fiber coupled allowing for easy distribution to further elements of the network where the short temporal profile of the photons is beneficial for impedance matching \cite{IonSemiConductorSteinerMeyer}. 
Compared to previous cavity-based atom-photon entanglements we achieved to our knowledge the yet shortest temporal shape of photons of $ 9.9(7) $\,ns extracted through the cavity by more than one order of magnitude. At the same time we maintained a high success rate of generating atom-photon entanglement per shot $ 2.5 \cdot 10^{-3}$  which is  comparable with the results of \cite{PhysRevLett.124.110501} who reported the yet highest measurement rate of atom-photon entanglement. 

In the future, we expect to increase the detection rate of entanglements of $ 62 \text{\,s}^{-1}  $ by an order of magnitude when approaching the maximum repetition rate allowed for by the ion initialisation time of 3.25\,$ \mu $s. Further improvements to ion localisation and locking quality of the cavity could yield another factor of five. Furthermore, the coherence time could be enhanced by shifting the atomic qubit to the clock transition of \element{171}Yb\element{+} using additional microwave pulses \cite{PhysRevA.76.052314}. A logical next step in the realization of a quantum network node would be the inclusion of a frequency conversion to telecom wavelength, which has been demonstrated recently for Ca$^+$ ions \cite{Bock2018,Krutyanskiy2019} and single Rubidium atoms \cite{Ikuta2018}. The principle of a conversion between 370\,nm  and 1314\,nm \cite{PhysRevApplied.7.024021} or 1580.3\,nm \cite{Kasture_2016} has been demonstrated already for weak laser fields and also conversion pathways for matching the trapped ion qubit to semiconductor quantum dots \cite{IonSemiConductorSteinerMeyer} or other emitters can be explored.

 

\section*{Methods}
\noindent {\bf Fiber Fabry-Perot cavity}\\
The fiber Fabry-Perot  cavity composes of two single-mode optical fibers, each having a concave mirror structure on its front facet with radii of curvature of $ R_1=\left(255\pm 16\right)$\,$\mu$m  and  $R_2=\left(304\pm34\right)$\,$\mu$m and reflectivities of $ T_1=500 $\,ppm and $ T_2=100 $\,ppm, respectively. From the measured finesse of $F=4700\pm700$ we estimate the losses of the mirrors to be $ L=(350\pm100)\text{\,ppm}$, in agreement with the manufacturer's expectation. Similar to our previous work on ultraviolet cavities \cite{doi:10.1063/1.5093551}, the losses have remained constant over time even under ultrahigh vacuum conditions. With a cavity length of $ 261\pm1 $\,$ \mu $m we calculate a mode waist ($ 1/e^2 $ intensity radius) of $ \omega_\text{cavity}= 4.1(2) $\,$\mu$m.  We stabilize the cavity length using a side-of-fringe locking technique using an auxiliary laser detuned by one free spectral range from the resonance of the atom. In order to avoid charging of the dielectric mirror surfaces due to ultraviolet light \cite{DielectricChargingUV} coupled into the cavity, we conduct the lock at ultralow light levels of less than $ 50 \text{\,pW} $. 
\newline
\\{\bf Pulsed excitation }\\
    We derive the excitation pulse from a mode-locked frequency-doubled Ti:sapphire laser running at 54\,MHz intra cavity repetition rate. This results in 18.5\,ns uncertainty of the pulse arrival time at the ion position. We synchronise parts of the experimental sequence to a trigger signal sent out with 50\,ps jitter simultaneously to the outcoupling of a laser pulse. The pulses  exhibit a spectral linewidth of $ \sim 90\text{\,GHz} $ which we filter using a cavity resonant to the ion excitation transition with linewidth of $ 2\pi\cdot 1.08(3)\text{\,GHz} $.
	This not only reduces unwanted excitations of the ion but also suppresses scattering of stray photons into the fiber-cavity detuned by 12.6\,GHz from the pulsed excitation frequency.
\newline
\\{\bf Detection efficiency }\\
We estimate the effective light-matter coupling rate $ g_\text{eff}=\sqrt{C_{0,\text{eff}}\cdot2\cdot \kappa \cdot \gamma}=2\pi\cdot\left(7.9\pm1.0\right) $\,MHz from the Purcell-enhanced spontaneous decay of the ion into a single mode of the resonator. The cavity linewidth $ \kappa$ dictates the decay constant of the intra-cavity electric field of $ \tau_\text{cavity} = \left(1.3\pm0.2\right)$\,ns, which we take into account for the decay time of the ion's excited state using two convolved exponential decays. With the Purcell-enhanced linewidth $ \Gamma^{'}=\Gamma\left(1+2C_{0, \text{eff}}\right)= 2\pi\cdot 21.58(45) $\,MHz and the atomic linewidth of $ 2\gamma=\Gamma=2\pi\cdot19.4 $\,MHz we calculate the effective cooperativity $ C_{0, \text{eff}}= 0.056(12)$. The effective value includes the effects of an imperfect locking of the cavity onto the atomic resonance wavelength and the imperfect localization of the ion due to residual (micro--)motion. From the effective cooperativity we calculate the probability of a photon emitted into the cavity mode $P_\text{c,eff}=\frac {2C_{0, \text{eff}}}{2C_{0, \text{eff}}+1}= (10.1\pm1.9)\% $. We determine the efficiency of detecting the photon on one of the single-photon counting modules to be $ P_\text{d,eff}=P_\text{c,eff}\cdot\eta_\text{ext}\cdot \epsilon \cdot \eta_\text{path} \cdot \eta_\text{Detector} = 3.2(9)\cdot10^{-3}$. Here, we use  the probability of the photon to be extracted through the desired low-reflectivity cavity mirror  $ \eta_\text{ext}=\frac{T_2}{T_1+T_2+L_{1+2}}=0.53(6) $, the mode matching efficiency into the single-mode optical fiber $ \epsilon=0.44(3) $, the quantum efficiency of the single-photon counters of $ \eta_\text{Detector}=0.215 $, and path efficiency $ \eta_\text{path}=0.65(11)$ due to optical fibers and further optical elements. The calculated value agrees with the measured detection efficiency of $ P_{\text{d,measured}} = 2.5 \cdot10^{-3}$. 
\newline
\\{\bf Lower bound of fidelity}\\
The lower bound of the entanglement fidelity can be estimated using the formula \cite{Blinov2004,Vasconcelos2020}

\begin{equation}
\begin{aligned}
F\geq{} & \frac{1}{2}(\rho_{\uparrow V,\uparrow V}+\rho_{\downarrow H,\downarrow H}-\sqrt{\rho_{\uparrow H,\uparrow H}\cdot\rho_{\downarrow V,\downarrow V}} \\
	& +\tilde{\rho}_{\uparrow V,\uparrow V}+\tilde{\rho}_{\downarrow H,\downarrow H}-\tilde{\rho}_{\uparrow H,\uparrow H}-\tilde{\rho}_{\downarrow V,\downarrow V}).
\end{aligned}
\end{equation}

With $ \ket{\uparrow} \equiv \ket{g^{+}} $ and $ \ket{\downarrow} \equiv \ket{0} $ the terms $ \rho_{\sigma\lambda,\sigma\lambda}=P(\sigma|\lambda)\cdot P(\lambda) $ are expressed in terms of measured probabilities. Here $ \lambda=\{H, V\} $ are the  photon states and $ \sigma=\{\uparrow,\downarrow \}$ the atom states. For the $ z $--basis the $ \rho_{\sigma\lambda,\sigma\lambda} $ can be directly extracted from Figure \ref{Entanglement}a whereas for the rotated basis the $\tilde{\rho} _{\sigma\lambda,\sigma\lambda} $ from Figure \ref{Entanglement}b. 
When corrected for the false readouts per sequence originating from dark counts of the cavity SPCs of $ 1.70(49)\cdot 10^{-6} $ \{$ 6.64(97)\cdot 10^{-6} $\} for H \{V\} we achieve a lower bound on the generated atom photon state fidelity of  $ F\geq90.7\pm1.5 $\,\%. We subtract half of the expected dark--counts on the bright and dark correlation side of the atom, since we expect an equal probability for a dark or bright ion.  
In the rotated basis two effects reduce the contrast of correlations: (1) The trade-off between timing precision of the microwave pulses and readout fidelity of the fluorescence based state detection of the ion. For lower magnetic field values we can increase the relative timing precision due to a reduced $ \omega_L $. However, falling below a certain magnetic field threshold reduces the fidelity of atomic state detection due to the formation of coherent dark states as described later in methods.
Considering all timing uncertainties including the post selection of emitted photons, we end up with $ \leq 0.05 \pi $ uncertainty of the atomic state in equatorial plane of the Bloch-sphere resulting in a reduction of contrast in the rotated bases of $ \leq 5 \% $.

\begin{figure*}
	\centering
	\includegraphics[width=0.8\textwidth]{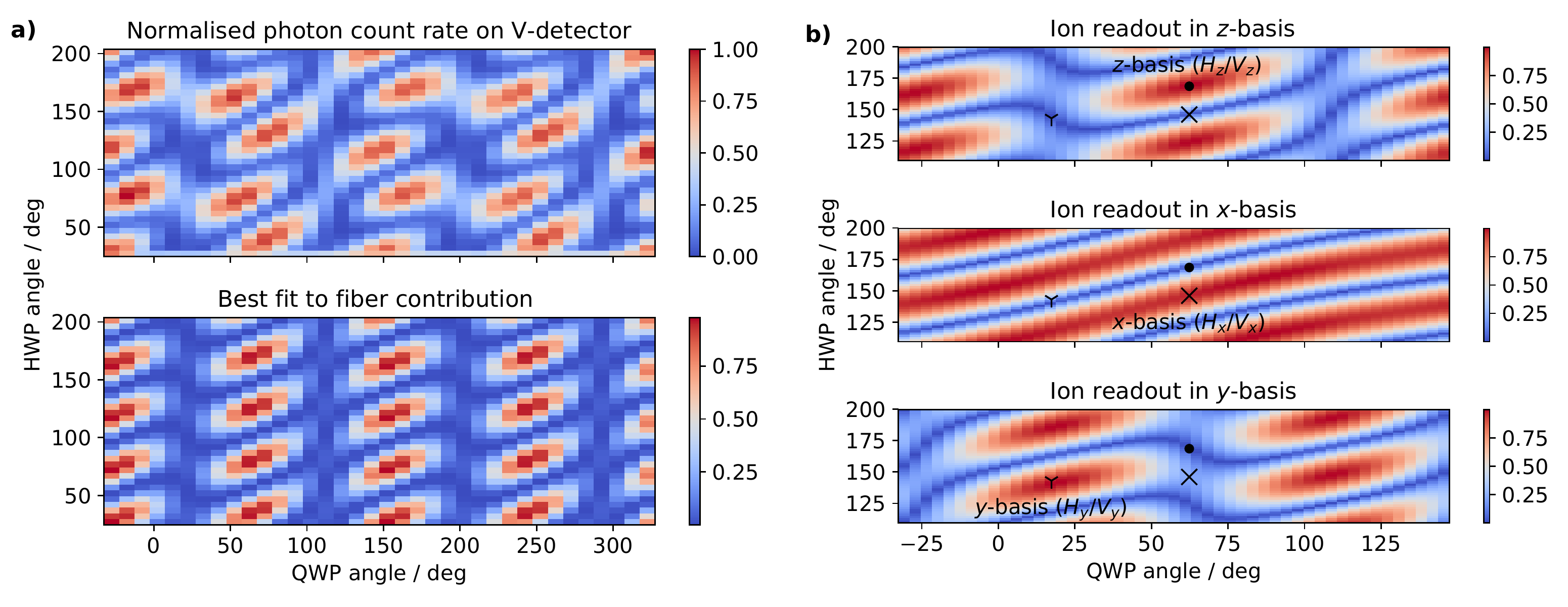}
	\caption{{\bf Selection of the photon read out basis} a) Characterisation of the photon detection setup using a reference laser as shown in Figure \ref{lvlschemeAndQubitManipulation}. The normalised count rate of reflected photons on the V--detector for different waveplate angles is shown (top). Fit of the data assuming the fiber to be an arbitrary polarization element (bottom). b) Calculated correlation heat-map for different atomic qubit basis using fiber parameters determined from the fit in a). The optimal waveplate settings of the photon readout basis $ \sigma_i $ used for the measurements in Figure \ref{Entanglement} are labelled accordingly with lower indices $ \text{H}_i/\text{V}_i;\text{ }i\in(x,y,z) $. }
	\label{wavePlateCavitySimulation}
\end{figure*}

(2) The projection of the ion to a superposition state of $ \ket{g^-} $ and $ \ket{g^+} $ is sensitive to magnetic field noise. From Figure  \ref{Coherence}b we estimate the reduction of contrast originating from magnetic field noise within the 57\,$ \mu $s of pulse sequence to be $ (5\pm3)$\,\%. 
\newline
\\{\bf Full quantum state tomography }\\
We reconstruct the density matrix from the joint expectation values $ S_{i,j} $ for 16 different combinations of Pauli basis $ \sigma_i \otimes \sigma_j $ with $ \sigma_{i/j} \in \{1,\sigma_x,\sigma_y,\sigma_z\}$. We make use of the symmetry of the off-diagonal elements for the $ x-, y- $ and $ z-$bases $  S_{m,n}=S_{n,m}$ for $m/n\in\{x,y,z\}$ and $m\neq n $. We extract the values from the measurements shown in Figure \ref{Entanglement} to compute:
\begin{equation}
\tilde{\rho}=\frac{1}{4}\sum_{i,j=0}^{3}S_{i,j} \sigma_i \otimes \sigma_j
\end{equation} 
Subsequently, we apply unitary transformations on the basis of each qubit in order to  post-hoc  maximise the overlap with the Bell state $ \frac{1}{\sqrt{2}}\left(\ket{V}\ket{g^+}-\ket{H}\ket{g^-}\right) $. Following \cite{PhysRevA.64.052312}, we perform a maximum likelihood estimation to ensure the density matrix $ \rho $ to be physical with the previously obtained matrix as starting point. We use this matrix to estimate the contribution of unwanted unitary rotations on the atom and photon basis and to compute an upper bound of the fidelity.
\newline
\\{\bf Magnetic field stability }\\
Our magnetic field noise is estimated to be smaller than 0.7\,mG with main noise contributions at 50\,Hz and 150\,Hz. We measure the magnetic field noise using a fluxgate sensor placed 10\,cm away from the ion's position but outside the stainless steel vacuum chamber. In order to reduce magnetic field noise the chamber is surrounded on 5 out of 6 sides with 3 layers of cobalt foil \footnote{MCL61}. 
\newline
\\{\bf Dark state repumping}\\
For magnetic fields lower than 1\,G ($ \approx 3 \text{\,MHz} $ splitting) we observe a drastic decrease of the ion fluorescence resulting from formation of coherent dark states of the \spectralLvl{2}{S}{1/2} , $ F=1 $  hyperfine levels \cite{PhysRevA.65.033413,PhysRevLett.80.940}. Decreasing the magnetic field into this range, however, is necessary to reduce the Larmor frequency and therefore improve timing synchronization. We are able to partly recover the fluorescence ($ \sim 50\% $) by using a second cooling laser with different polarisation through a different beam path destroying the dark states.  For the ion state readout based on free-space fluorescence detection, this second laser beam results in an increase of dark counts while at the same time the bright counts are reduced by a factor of two by dark states. Both effects reduce the contrast of the fluorescence based atomic readout by  $ \lesssim 10$\,\%. 
\newline
\\{\bf Photonic basis}\\
The photonic readout basis is defined by a half-- and a quarter--wave plate (HWP \& QWP) which rotate the basis of the polarisation qubit. A projective measurement is achieved by a polarising beam splitter (PBS) and two single photon counters (SPCs) on each exit path of the PBS respectively, detecting horizontal (H) or vertical (V) polarised photons (see Figure \ref{lvlschemeAndQubitManipulation}a). The two waveplates can select any basis of the photons extracted through the fiber-cavity to be projected as (H/V) on the two SPCs. For a defined selection of a basis, knowledge about the influence of the fiber on the polarisation is required.  We characterise the photon path using a weak laser ($ \sim50 $\,pW) coupled through the PBS into the fiber. Since the laser frequency is off-resonant by a few GHz with respect to the cavity, it gets reflected on the first fiber mirror and we detect the reflected light on the V-SPC. Repeating the measurement for different waveplate rotation angles leads to a heat map as shown in Figure \ref{wavePlateCavitySimulation}. Using the Jones--formalism we are able to do a best fit to the acquired data and find the polarization retardation of the fiber as well as the rotation offset of the fast axis of our waveplates.
As an example, for selecting the $\sigma_z$--basis, fiber, QWP and HWP should act in total as a quarter-wave plate mapping a circular polarised photon to a linear polarised (and vise versa). For the reflection of the initial H-polarised reference light this results in maximal counts on the V-SPC. 
When selecting bases orthogonal to $ \sigma_z $ we found that defining the $ \sigma_{x/y} $--basis as $ \sigma_k=\{\frac{1}{\sqrt{2}}\left(\ket{\sigma^+}+e^{i\phi_k}\ket{\sigma^-}\right),\frac{1}{\sqrt{2}}\left(\ket{\sigma^+}+e^{i\phi_k+\pi}\ket{\sigma^-}\right)\},\text{ } k\in(x,y)$ with $ \phi_x=\pi/4$ and $ \phi_y=-\pi/4 $ (same holds then for the ion basis) results in easily accessible waveplate angles as shown in Figure \ref{wavePlateCavitySimulation}. 
\newline
\section{Data Availability}
The data and code that support the findings of this study are available from the corresponding author upon reasonable request.
\newline
\section{ACKNOWLEDGEMENTS}
We thank T.G. Ballance, K. Kluge, and H.M. Meyer for early contributions to this work and R. Berner for experimental assistance.  This work has been funded by the Alexander-von-Humboldt Stiftung, DFG (SFB/TR 185 project A2), BMBF (FaResQ and Q.Link.X), and the Deutsche Forschungsgemeinschaft (DFG, German Research Foundation) under Germany's Excellence Strategy – Cluster of Excellence Matter and Light for Quantum Computing (ML4Q) EXC 2004/1 – 390534769.
\newline
\section{AUTHOR CONTRIBUTIONS}
All authors contributed to the design, development and characterisation of the experiment. P.K. conducted the measurements and the data analysis, M.B.  manufactured and characterised the fiber cavity setup.  P.K. and M.K.  wrote the paper. The project was conceived and supervised by M.K.
\newline
\section{COMPETING INTERESTS}
The authors declare no competing interests.
%
%
%

%


\begin{thebibliography}{43}%
	\makeatletter
	\providecommand \@ifxundefined [1]{%
		\@ifx{#1\undefined}
	}%
	\providecommand \@ifnum [1]{%
		\ifnum #1\expandafter \@firstoftwo
		\else \expandafter \@secondoftwo
		\fi
	}%
	\providecommand \@ifx [1]{%
		\ifx #1\expandafter \@firstoftwo
		\else \expandafter \@secondoftwo
		\fi
	}%
	\providecommand \natexlab [1]{#1}%
	\providecommand \enquote  [1]{``#1''}%
	\providecommand \bibnamefont  [1]{#1}%
	\providecommand \bibfnamefont [1]{#1}%
	\providecommand \citenamefont [1]{#1}%
	\providecommand \href@noop [0]{\@secondoftwo}%
	\providecommand \href [0]{\begingroup \@sanitize@url \@href}%
	\providecommand \@href[1]{\@@startlink{#1}\@@href}%
	\providecommand \@@href[1]{\endgroup#1\@@endlink}%
	\providecommand \@sanitize@url [0]{\catcode `\\12\catcode `\$12\catcode
		`\&12\catcode `\#12\catcode `\^12\catcode `\_12\catcode `\%12\relax}%
	\providecommand \@@startlink[1]{}%
	\providecommand \@@endlink[0]{}%
	\providecommand \url  [0]{\begingroup\@sanitize@url \@url }%
	\providecommand \@url [1]{\endgroup\@href {#1}{\urlprefix }}%
	\providecommand \urlprefix  [0]{URL }%
	\providecommand \Eprint [0]{\href }%
	\providecommand \doibase [0]{http://dx.doi.org/}%
	\providecommand \selectlanguage [0]{\@gobble}%
	\providecommand \bibinfo  [0]{\@secondoftwo}%
	\providecommand \bibfield  [0]{\@secondoftwo}%
	\providecommand \translation [1]{[#1]}%
	\providecommand \BibitemOpen [0]{}%
	\providecommand \bibitemStop [0]{}%
	\providecommand \bibitemNoStop [0]{.\EOS\space}%
	\providecommand \EOS [0]{\spacefactor3000\relax}%
	\providecommand \BibitemShut  [1]{\csname bibitem#1\endcsname}%
	\let\auto@bib@innerbib\@empty
	\bibitem [{\citenamefont {Kimble}(2008)}]{QuantumInternet}%
	\BibitemOpen
	\bibfield  {author} {\bibinfo {author} {\bibfnamefont {H.~J.}\ \bibnamefont
			{Kimble}},\ }\href {\doibase 10.1038/nature07127} {\bibfield  {journal}
		{\bibinfo  {journal} {Nature}\ }\textbf {\bibinfo {volume} {453}},\ \bibinfo
		{pages} {1023} (\bibinfo {year} {2008})}\BibitemShut {NoStop}%
	\bibitem [{\citenamefont {Degen}\ \emph {et~al.}(2017)\citenamefont {Degen},
		\citenamefont {Reinhard},\ and\ \citenamefont
		{Cappellaro}}]{RevModPhys.89.035002}%
	\BibitemOpen
	\bibfield  {author} {\bibinfo {author} {\bibfnamefont {C.~L.}\ \bibnamefont
			{Degen}}, \bibinfo {author} {\bibfnamefont {F.}~\bibnamefont {Reinhard}}, \
		and\ \bibinfo {author} {\bibfnamefont {P.}~\bibnamefont {Cappellaro}},\
	}\href {\doibase 10.1103/RevModPhys.89.035002} {\bibfield  {journal}
		{\bibinfo  {journal} {Rev. Mod. Phys.}\ }\textbf {\bibinfo {volume} {89}},\
		\bibinfo {pages} {035002} (\bibinfo {year} {2017})}\BibitemShut {NoStop}%
	\bibitem [{\citenamefont {Wootters}\ and\ \citenamefont
		{Zurek}(1982)}]{NonCloningTheorem}%
	\BibitemOpen
	\bibfield  {author} {\bibinfo {author} {\bibfnamefont {W.~K.}\ \bibnamefont
			{Wootters}}\ and\ \bibinfo {author} {\bibfnamefont {W.~H.}\ \bibnamefont
			{Zurek}},\ }\href@noop {} {\bibfield  {journal} {\bibinfo  {journal}
			{Nature}\ }\textbf {\bibinfo {volume} {299}},\ \bibinfo {pages} {802}
		(\bibinfo {year} {1982})}\BibitemShut {NoStop}%
	\bibitem [{\citenamefont {Shor}\ and\ \citenamefont
		{Preskill}(2000)}]{QuantumKeyDistrBB84}%
	\BibitemOpen
	\bibfield  {author} {\bibinfo {author} {\bibfnamefont {P.~W.}\ \bibnamefont
			{Shor}}\ and\ \bibinfo {author} {\bibfnamefont {J.}~\bibnamefont
			{Preskill}},\ }\href@noop {} {\bibfield  {journal} {\bibinfo  {journal}
			{Phys. Rev. Lett.}\ }\textbf {\bibinfo {volume} {85}},\ \bibinfo {pages}
		{441} (\bibinfo {year} {2000})}\BibitemShut {NoStop}%
	\bibitem [{\citenamefont {Blinov}\ \emph {et~al.}(2004)\citenamefont {Blinov},
		\citenamefont {Moehring}, \citenamefont {Duan},\ and\ \citenamefont
		{Monroe}}]{Blinov2004}%
	\BibitemOpen
	\bibfield  {author} {\bibinfo {author} {\bibfnamefont {B.~B.}\ \bibnamefont
			{Blinov}}, \bibinfo {author} {\bibfnamefont {D.~L.}\ \bibnamefont
			{Moehring}}, \bibinfo {author} {\bibfnamefont {L.-M.}\ \bibnamefont {Duan}},
		\ and\ \bibinfo {author} {\bibfnamefont {C.}~\bibnamefont {Monroe}},\ }\href
	{\doibase 10.1038/nature02377} {\bibfield  {journal} {\bibinfo  {journal}
			{Nature}\ }\textbf {\bibinfo {volume} {428}},\ \bibinfo {pages} {153}
		(\bibinfo {year} {2004})}\BibitemShut {NoStop}%
	\bibitem [{\citenamefont {Stute}\ \emph {et~al.}(2012)\citenamefont {Stute},
		\citenamefont {Casabone}, \citenamefont {Schindler}, \citenamefont {Monz},
		\citenamefont {Schmidt}, \citenamefont {Brandstätter}, \citenamefont
		{Northup},\ and\ \citenamefont {Blatt}}]{2012_fastest_spin_photon}%
	\BibitemOpen
	\bibfield  {author} {\bibinfo {author} {\bibfnamefont {A.}~\bibnamefont
			{Stute}}, \bibinfo {author} {\bibfnamefont {B.}~\bibnamefont {Casabone}},
		\bibinfo {author} {\bibfnamefont {P.}~\bibnamefont {Schindler}}, \bibinfo
		{author} {\bibfnamefont {T.}~\bibnamefont {Monz}}, \bibinfo {author}
		{\bibfnamefont {P.~O.}\ \bibnamefont {Schmidt}}, \bibinfo {author}
		{\bibfnamefont {B.}~\bibnamefont {Brandstätter}}, \bibinfo {author}
		{\bibfnamefont {T.~E.}\ \bibnamefont {Northup}}, \ and\ \bibinfo {author}
		{\bibfnamefont {R.}~\bibnamefont {Blatt}},\ }\href {\doibase
		10.1038/nature11120} {\bibfield  {journal} {\bibinfo  {journal} {Nature}\
		}\textbf {\bibinfo {volume} {485}},\ \bibinfo {pages} {482} (\bibinfo {year}
		{2012})}\BibitemShut {NoStop}%
	\bibitem [{\citenamefont {Bock}\ \emph {et~al.}(2018)\citenamefont {Bock},
		\citenamefont {Eich}, \citenamefont {Kucera}, \citenamefont {Kreis},
		\citenamefont {Lenhard}, \citenamefont {Becher},\ and\ \citenamefont
		{Eschner}}]{Bock2018}%
	\BibitemOpen
	\bibfield  {author} {\bibinfo {author} {\bibfnamefont {M.}~\bibnamefont
			{Bock}}, \bibinfo {author} {\bibfnamefont {P.}~\bibnamefont {Eich}}, \bibinfo
		{author} {\bibfnamefont {S.}~\bibnamefont {Kucera}}, \bibinfo {author}
		{\bibfnamefont {M.}~\bibnamefont {Kreis}}, \bibinfo {author} {\bibfnamefont
			{A.}~\bibnamefont {Lenhard}}, \bibinfo {author} {\bibfnamefont
			{C.}~\bibnamefont {Becher}}, \ and\ \bibinfo {author} {\bibfnamefont
			{J.}~\bibnamefont {Eschner}},\ }\href {\doibase 10.1038/s41467-018-04341-2}
	{\bibfield  {journal} {\bibinfo  {journal} {Nature Communications}\ }\textbf
		{\bibinfo {volume} {9}},\ \bibinfo {pages} {1998} (\bibinfo {year}
		{2018})}\BibitemShut {NoStop}%
	\bibitem [{\citenamefont {Stute}\ \emph {et~al.}(2013)\citenamefont {Stute},
		\citenamefont {Casabone}, \citenamefont {Brandst{\"a}tter}, \citenamefont
		{Friebe}, \citenamefont {Northup},\ and\ \citenamefont {Blatt}}]{Stute2013}%
	\BibitemOpen
	\bibfield  {author} {\bibinfo {author} {\bibfnamefont {A.}~\bibnamefont
			{Stute}}, \bibinfo {author} {\bibfnamefont {B.}~\bibnamefont {Casabone}},
		\bibinfo {author} {\bibfnamefont {B.}~\bibnamefont {Brandst{\"a}tter}},
		\bibinfo {author} {\bibfnamefont {K.}~\bibnamefont {Friebe}}, \bibinfo
		{author} {\bibfnamefont {T.~E.}\ \bibnamefont {Northup}}, \ and\ \bibinfo
		{author} {\bibfnamefont {R.}~\bibnamefont {Blatt}},\ }\href {\doibase
		10.1038/nphoton.2012.358} {\bibfield  {journal} {\bibinfo  {journal} {Nature
				Photonics}\ }\textbf {\bibinfo {volume} {7}},\ \bibinfo {pages} {219}
		(\bibinfo {year} {2013})}\BibitemShut {NoStop}%
	\bibitem [{\citenamefont {Wilk}\ \emph {et~al.}(2007)\citenamefont {Wilk},
		\citenamefont {Webster}, \citenamefont {Kuhn},\ and\ \citenamefont
		{Rempe}}]{Wilk488}%
	\BibitemOpen
	\bibfield  {author} {\bibinfo {author} {\bibfnamefont {T.}~\bibnamefont
			{Wilk}}, \bibinfo {author} {\bibfnamefont {S.~C.}\ \bibnamefont {Webster}},
		\bibinfo {author} {\bibfnamefont {A.}~\bibnamefont {Kuhn}}, \ and\ \bibinfo
		{author} {\bibfnamefont {G.}~\bibnamefont {Rempe}},\ }\href {\doibase
		10.1126/science.1143835} {\bibfield  {journal} {\bibinfo  {journal}
			{Science}\ }\textbf {\bibinfo {volume} {317}},\ \bibinfo {pages} {488}
		(\bibinfo {year} {2007})}\BibitemShut {NoStop}%
	\bibitem [{\citenamefont {Volz}\ \emph {et~al.}(2006)\citenamefont {Volz},
		\citenamefont {Weber}, \citenamefont {Schlenk}, \citenamefont {Rosenfeld},
		\citenamefont {Vrana}, \citenamefont {Saucke}, \citenamefont {Kurtsiefer},\
		and\ \citenamefont {Weinfurter}}]{PhysRevLett.96.030404}%
	\BibitemOpen
	\bibfield  {author} {\bibinfo {author} {\bibfnamefont {J.}~\bibnamefont
			{Volz}}, \bibinfo {author} {\bibfnamefont {M.}~\bibnamefont {Weber}},
		\bibinfo {author} {\bibfnamefont {D.}~\bibnamefont {Schlenk}}, \bibinfo
		{author} {\bibfnamefont {W.}~\bibnamefont {Rosenfeld}}, \bibinfo {author}
		{\bibfnamefont {J.}~\bibnamefont {Vrana}}, \bibinfo {author} {\bibfnamefont
			{K.}~\bibnamefont {Saucke}}, \bibinfo {author} {\bibfnamefont
			{C.}~\bibnamefont {Kurtsiefer}}, \ and\ \bibinfo {author} {\bibfnamefont
			{H.}~\bibnamefont {Weinfurter}},\ }\href {\doibase
		10.1103/PhysRevLett.96.030404} {\bibfield  {journal} {\bibinfo  {journal}
			{Phys. Rev. Lett.}\ }\textbf {\bibinfo {volume} {96}},\ \bibinfo {pages}
		{030404} (\bibinfo {year} {2006})}\BibitemShut {NoStop}%
	\bibitem [{\citenamefont {Togan}\ \emph {et~al.}(2010)\citenamefont {Togan},
		\citenamefont {Chu}, \citenamefont {Trifonov}, \citenamefont {Jiang},
		\citenamefont {Maze}, \citenamefont {Childress}, \citenamefont {Dutt},
		\citenamefont {Sørensen}, \citenamefont {Hemmer}, \citenamefont {Zibrov},\
		and\ \citenamefont {Lukin}}]{Togan2010}%
	\BibitemOpen
	\bibfield  {author} {\bibinfo {author} {\bibfnamefont {E.}~\bibnamefont
			{Togan}}, \bibinfo {author} {\bibfnamefont {Y.}~\bibnamefont {Chu}}, \bibinfo
		{author} {\bibfnamefont {A.}~\bibnamefont {Trifonov}}, \bibinfo {author}
		{\bibfnamefont {L.}~\bibnamefont {Jiang}}, \bibinfo {author} {\bibfnamefont
			{J.}~\bibnamefont {Maze}}, \bibinfo {author} {\bibfnamefont {L.}~\bibnamefont
			{Childress}}, \bibinfo {author} {\bibfnamefont {G.}~\bibnamefont {Dutt}},
		\bibinfo {author} {\bibfnamefont {A.}~\bibnamefont {Sørensen}}, \bibinfo
		{author} {\bibfnamefont {P.}~\bibnamefont {Hemmer}}, \bibinfo {author}
		{\bibfnamefont {A.}~\bibnamefont {Zibrov}}, \ and\ \bibinfo {author}
		{\bibfnamefont {M.}~\bibnamefont {Lukin}},\ }\href {\doibase
		10.1038/nature09256} {\bibfield  {journal} {\bibinfo  {journal} {Nature}\
		}\textbf {\bibinfo {volume} {466}},\ \bibinfo {pages} {730} (\bibinfo {year}
		{2010})}\BibitemShut {NoStop}%
	\bibitem [{\citenamefont {Nguyen}\ \emph {et~al.}(2019)\citenamefont {Nguyen},
		\citenamefont {Sukachev}, \citenamefont {Bhaskar}, \citenamefont {Machielse},
		\citenamefont {Levonian}, \citenamefont {Knall}, \citenamefont {Stroganov},
		\citenamefont {Riedinger}, \citenamefont {Park}, \citenamefont
		{Lon\ifmmode~\check{c}\else \v{c}\fi{}ar},\ and\ \citenamefont
		{Lukin}}]{PhysRevLett.123.183602}%
	\BibitemOpen
	\bibfield  {author} {\bibinfo {author} {\bibfnamefont {C.~T.}\ \bibnamefont
			{Nguyen}}, \bibinfo {author} {\bibfnamefont {D.~D.}\ \bibnamefont
			{Sukachev}}, \bibinfo {author} {\bibfnamefont {M.~K.}\ \bibnamefont
			{Bhaskar}}, \bibinfo {author} {\bibfnamefont {B.}~\bibnamefont {Machielse}},
		\bibinfo {author} {\bibfnamefont {D.~S.}\ \bibnamefont {Levonian}}, \bibinfo
		{author} {\bibfnamefont {E.~N.}\ \bibnamefont {Knall}}, \bibinfo {author}
		{\bibfnamefont {P.}~\bibnamefont {Stroganov}}, \bibinfo {author}
		{\bibfnamefont {R.}~\bibnamefont {Riedinger}}, \bibinfo {author}
		{\bibfnamefont {H.}~\bibnamefont {Park}}, \bibinfo {author} {\bibfnamefont
			{M.}~\bibnamefont {Lon\ifmmode~\check{c}\else \v{c}\fi{}ar}}, \ and\ \bibinfo
		{author} {\bibfnamefont {M.~D.}\ \bibnamefont {Lukin}},\ }\href {\doibase
		10.1103/PhysRevLett.123.183602} {\bibfield  {journal} {\bibinfo  {journal}
			{Phys. Rev. Lett.}\ }\textbf {\bibinfo {volume} {123}},\ \bibinfo {pages}
		{183602} (\bibinfo {year} {2019})}\BibitemShut {NoStop}%
	\bibitem [{\citenamefont {De~Greve}\ \emph {et~al.}(2012)\citenamefont
		{De~Greve}, \citenamefont {Yu}, \citenamefont {McMahon}, \citenamefont
		{Pelc}, \citenamefont {Natarajan}, \citenamefont {Kim}, \citenamefont {Abe},
		\citenamefont {Maier}, \citenamefont {Schneider}, \citenamefont {Kamp},
		\citenamefont {H{\"o}fling}, \citenamefont {Hadfield}, \citenamefont
		{Forchel}, \citenamefont {Fejer},\ and\ \citenamefont
		{Yamamoto}}]{DeGreve2012}%
	\BibitemOpen
	\bibfield  {author} {\bibinfo {author} {\bibfnamefont {K.}~\bibnamefont
			{De~Greve}}, \bibinfo {author} {\bibfnamefont {L.}~\bibnamefont {Yu}},
		\bibinfo {author} {\bibfnamefont {P.~L.}\ \bibnamefont {McMahon}}, \bibinfo
		{author} {\bibfnamefont {J.~S.}\ \bibnamefont {Pelc}}, \bibinfo {author}
		{\bibfnamefont {C.~M.}\ \bibnamefont {Natarajan}}, \bibinfo {author}
		{\bibfnamefont {N.~Y.}\ \bibnamefont {Kim}}, \bibinfo {author} {\bibfnamefont
			{E.}~\bibnamefont {Abe}}, \bibinfo {author} {\bibfnamefont {S.}~\bibnamefont
			{Maier}}, \bibinfo {author} {\bibfnamefont {C.}~\bibnamefont {Schneider}},
		\bibinfo {author} {\bibfnamefont {M.}~\bibnamefont {Kamp}}, \bibinfo {author}
		{\bibfnamefont {S.}~\bibnamefont {H{\"o}fling}}, \bibinfo {author}
		{\bibfnamefont {R.~H.}\ \bibnamefont {Hadfield}}, \bibinfo {author}
		{\bibfnamefont {A.}~\bibnamefont {Forchel}}, \bibinfo {author} {\bibfnamefont
			{M.~M.}\ \bibnamefont {Fejer}}, \ and\ \bibinfo {author} {\bibfnamefont
			{Y.}~\bibnamefont {Yamamoto}},\ }\href {\doibase 10.1038/nature11577}
	{\bibfield  {journal} {\bibinfo  {journal} {Nature}\ }\textbf {\bibinfo
			{volume} {491}},\ \bibinfo {pages} {421} (\bibinfo {year}
		{2012})}\BibitemShut {NoStop}%
	\bibitem [{\citenamefont {Gao}\ \emph {et~al.}(2012)\citenamefont {Gao},
		\citenamefont {Fallahi}, \citenamefont {Togan}, \citenamefont
		{Miguel-Sanchez},\ and\ \citenamefont {Imamoglu}}]{Gao2012}%
	\BibitemOpen
	\bibfield  {author} {\bibinfo {author} {\bibfnamefont {W.~B.}\ \bibnamefont
			{Gao}}, \bibinfo {author} {\bibfnamefont {P.}~\bibnamefont {Fallahi}},
		\bibinfo {author} {\bibfnamefont {E.}~\bibnamefont {Togan}}, \bibinfo
		{author} {\bibfnamefont {J.}~\bibnamefont {Miguel-Sanchez}}, \ and\ \bibinfo
		{author} {\bibfnamefont {A.}~\bibnamefont {Imamoglu}},\ }\href {\doibase
		10.1038/nature11573} {\bibfield  {journal} {\bibinfo  {journal} {Nature}\
		}\textbf {\bibinfo {volume} {491}},\ \bibinfo {pages} {426} (\bibinfo {year}
		{2012})}\BibitemShut {NoStop}%
	\bibitem [{\citenamefont {Stephenson}\ \emph {et~al.}(2020)\citenamefont
		{Stephenson}, \citenamefont {Nadlinger}, \citenamefont {Nichol},
		\citenamefont {An}, \citenamefont {Drmota}, \citenamefont {Ballance},
		\citenamefont {Thirumalai}, \citenamefont {Goodwin}, \citenamefont {Lucas},\
		and\ \citenamefont {Ballance}}]{PhysRevLett.124.110501}%
	\BibitemOpen
	\bibfield  {author} {\bibinfo {author} {\bibfnamefont {L.~J.}\ \bibnamefont
			{Stephenson}}, \bibinfo {author} {\bibfnamefont {D.~P.}\ \bibnamefont
			{Nadlinger}}, \bibinfo {author} {\bibfnamefont {B.~C.}\ \bibnamefont
			{Nichol}}, \bibinfo {author} {\bibfnamefont {S.}~\bibnamefont {An}}, \bibinfo
		{author} {\bibfnamefont {P.}~\bibnamefont {Drmota}}, \bibinfo {author}
		{\bibfnamefont {T.~G.}\ \bibnamefont {Ballance}}, \bibinfo {author}
		{\bibfnamefont {K.}~\bibnamefont {Thirumalai}}, \bibinfo {author}
		{\bibfnamefont {J.~F.}\ \bibnamefont {Goodwin}}, \bibinfo {author}
		{\bibfnamefont {D.~M.}\ \bibnamefont {Lucas}}, \ and\ \bibinfo {author}
		{\bibfnamefont {C.~J.}\ \bibnamefont {Ballance}},\ }\href {\doibase
		10.1103/PhysRevLett.124.110501} {\bibfield  {journal} {\bibinfo  {journal}
			{Phys. Rev. Lett.}\ }\textbf {\bibinfo {volume} {124}},\ \bibinfo {pages}
		{110501} (\bibinfo {year} {2020})}\BibitemShut {NoStop}%
	\bibitem [{\citenamefont {Almendros}\ \emph {et~al.}(2009)\citenamefont
		{Almendros}, \citenamefont {Huwer}, \citenamefont {Piro}, \citenamefont
		{Rohde}, \citenamefont {Schuck}, \citenamefont {Hennrich}, \citenamefont
		{Dubin},\ and\ \citenamefont {Eschner}}]{PhysRevLett.103.213601}%
	\BibitemOpen
	\bibfield  {author} {\bibinfo {author} {\bibfnamefont {M.}~\bibnamefont
			{Almendros}}, \bibinfo {author} {\bibfnamefont {J.}~\bibnamefont {Huwer}},
		\bibinfo {author} {\bibfnamefont {N.}~\bibnamefont {Piro}}, \bibinfo {author}
		{\bibfnamefont {F.}~\bibnamefont {Rohde}}, \bibinfo {author} {\bibfnamefont
			{C.}~\bibnamefont {Schuck}}, \bibinfo {author} {\bibfnamefont
			{M.}~\bibnamefont {Hennrich}}, \bibinfo {author} {\bibfnamefont
			{F.}~\bibnamefont {Dubin}}, \ and\ \bibinfo {author} {\bibfnamefont
			{J.}~\bibnamefont {Eschner}},\ }\href {\doibase
		10.1103/PhysRevLett.103.213601} {\bibfield  {journal} {\bibinfo  {journal}
			{Phys. Rev. Lett.}\ }\textbf {\bibinfo {volume} {103}},\ \bibinfo {pages}
		{213601} (\bibinfo {year} {2009})}\BibitemShut {NoStop}%
	\bibitem [{\citenamefont {Gerber}\ \emph {et~al.}(2009)\citenamefont {Gerber},
		\citenamefont {Rotter}, \citenamefont {Hennrich}, \citenamefont {Blatt},
		\citenamefont {Rohde}, \citenamefont {Schuck}, \citenamefont {Almendros},
		\citenamefont {Gehr}, \citenamefont {Dubin},\ and\ \citenamefont
		{Eschner}}]{Gerber_2009}%
	\BibitemOpen
	\bibfield  {author} {\bibinfo {author} {\bibfnamefont {S.}~\bibnamefont
			{Gerber}}, \bibinfo {author} {\bibfnamefont {D.}~\bibnamefont {Rotter}},
		\bibinfo {author} {\bibfnamefont {M.}~\bibnamefont {Hennrich}}, \bibinfo
		{author} {\bibfnamefont {R.}~\bibnamefont {Blatt}}, \bibinfo {author}
		{\bibfnamefont {F.}~\bibnamefont {Rohde}}, \bibinfo {author} {\bibfnamefont
			{C.}~\bibnamefont {Schuck}}, \bibinfo {author} {\bibfnamefont
			{M.}~\bibnamefont {Almendros}}, \bibinfo {author} {\bibfnamefont
			{R.}~\bibnamefont {Gehr}}, \bibinfo {author} {\bibfnamefont {F.}~\bibnamefont
			{Dubin}}, \ and\ \bibinfo {author} {\bibfnamefont {J.}~\bibnamefont
			{Eschner}},\ }\href {\doibase 10.1088/1367-2630/11/1/013032} {\bibfield
		{journal} {\bibinfo  {journal} {New Journal of Physics}\ }\textbf {\bibinfo
			{volume} {11}},\ \bibinfo {pages} {013032} (\bibinfo {year}
		{2009})}\BibitemShut {NoStop}%
	\bibitem [{\citenamefont {Krutyanskiy}\ \emph {et~al.}(2019)\citenamefont
		{Krutyanskiy}, \citenamefont {Meraner}, \citenamefont {Schupp}, \citenamefont
		{Krcmarsky}, \citenamefont {Hainzer},\ and\ \citenamefont
		{Lanyon}}]{Krutyanskiy2019}%
	\BibitemOpen
	\bibfield  {author} {\bibinfo {author} {\bibfnamefont {V.}~\bibnamefont
			{Krutyanskiy}}, \bibinfo {author} {\bibfnamefont {M.}~\bibnamefont
			{Meraner}}, \bibinfo {author} {\bibfnamefont {J.}~\bibnamefont {Schupp}},
		\bibinfo {author} {\bibfnamefont {V.}~\bibnamefont {Krcmarsky}}, \bibinfo
		{author} {\bibfnamefont {H.}~\bibnamefont {Hainzer}}, \ and\ \bibinfo
		{author} {\bibfnamefont {B.~P.}\ \bibnamefont {Lanyon}},\ }\href {\doibase
		10.1038/s41534-019-0186-3} {\bibfield  {journal} {\bibinfo  {journal} {npj
				Quantum Information}\ }\textbf {\bibinfo {volume} {5}},\ \bibinfo {pages}
		{72} (\bibinfo {year} {2019})}\BibitemShut {NoStop}%
	\bibitem [{\citenamefont {Steiner}\ \emph {et~al.}(2013)\citenamefont
		{Steiner}, \citenamefont {Meyer}, \citenamefont {Deutsch}, \citenamefont
		{Reichel},\ and\ \citenamefont {K\"ohl}}]{PhysRevLett.110.043003}%
	\BibitemOpen
	\bibfield  {author} {\bibinfo {author} {\bibfnamefont {M.}~\bibnamefont
			{Steiner}}, \bibinfo {author} {\bibfnamefont {H.~M.}\ \bibnamefont {Meyer}},
		\bibinfo {author} {\bibfnamefont {C.}~\bibnamefont {Deutsch}}, \bibinfo
		{author} {\bibfnamefont {J.}~\bibnamefont {Reichel}}, \ and\ \bibinfo
		{author} {\bibfnamefont {M.}~\bibnamefont {K\"ohl}},\ }\href {\doibase
		10.1103/PhysRevLett.110.043003} {\bibfield  {journal} {\bibinfo  {journal}
			{Phys. Rev. Lett.}\ }\textbf {\bibinfo {volume} {110}},\ \bibinfo {pages}
		{043003} (\bibinfo {year} {2013})}\BibitemShut {NoStop}%
	\bibitem [{\citenamefont {Hunger}\ \emph {et~al.}(2010)\citenamefont {Hunger},
		\citenamefont {Steinmetz}, \citenamefont {Colombe}, \citenamefont {Deutsch},
		\citenamefont {Hänsch},\ and\ \citenamefont {Reichel}}]{Hunger_2010}%
	\BibitemOpen
	\bibfield  {author} {\bibinfo {author} {\bibfnamefont {D.}~\bibnamefont
			{Hunger}}, \bibinfo {author} {\bibfnamefont {T.}~\bibnamefont {Steinmetz}},
		\bibinfo {author} {\bibfnamefont {Y.}~\bibnamefont {Colombe}}, \bibinfo
		{author} {\bibfnamefont {C.}~\bibnamefont {Deutsch}}, \bibinfo {author}
		{\bibfnamefont {T.~W.}\ \bibnamefont {Hänsch}}, \ and\ \bibinfo {author}
		{\bibfnamefont {J.}~\bibnamefont {Reichel}},\ }\href {\doibase
		10.1088/1367-2630/12/6/065038} {\bibfield  {journal} {\bibinfo  {journal}
			{New Journal of Physics}\ }\textbf {\bibinfo {volume} {12}},\ \bibinfo
		{pages} {065038} (\bibinfo {year} {2010})}\BibitemShut {NoStop}%
	\bibitem [{\citenamefont {Northup}\ and\ \citenamefont
		{Blatt}(2014)}]{Northup2014}%
	\BibitemOpen
	\bibfield  {author} {\bibinfo {author} {\bibfnamefont {T.~E.}\ \bibnamefont
			{Northup}}\ and\ \bibinfo {author} {\bibfnamefont {R.}~\bibnamefont
			{Blatt}},\ }\href {\doibase 10.1038/nphoton.2014.53} {\bibfield  {journal}
		{\bibinfo  {journal} {Nature Photonics}\ }\textbf {\bibinfo {volume} {8}},\
		\bibinfo {pages} {356} (\bibinfo {year} {2014})}\BibitemShut {NoStop}%
	\bibitem [{\citenamefont {Gallego}\ \emph {et~al.}(2018)\citenamefont
		{Gallego}, \citenamefont {Alt}, \citenamefont {Macha}, \citenamefont
		{Martinez-Dorantes}, \citenamefont {Pandey},\ and\ \citenamefont
		{Meschede}}]{PhysRevLett.121.173603}%
	\BibitemOpen
	\bibfield  {author} {\bibinfo {author} {\bibfnamefont {J.}~\bibnamefont
			{Gallego}}, \bibinfo {author} {\bibfnamefont {W.}~\bibnamefont {Alt}},
		\bibinfo {author} {\bibfnamefont {T.}~\bibnamefont {Macha}}, \bibinfo
		{author} {\bibfnamefont {M.}~\bibnamefont {Martinez-Dorantes}}, \bibinfo
		{author} {\bibfnamefont {D.}~\bibnamefont {Pandey}}, \ and\ \bibinfo {author}
		{\bibfnamefont {D.}~\bibnamefont {Meschede}},\ }\href {\doibase
		10.1103/PhysRevLett.121.173603} {\bibfield  {journal} {\bibinfo  {journal}
			{Phys. Rev. Lett.}\ }\textbf {\bibinfo {volume} {121}},\ \bibinfo {pages}
		{173603} (\bibinfo {year} {2018})}\BibitemShut {NoStop}%
	\bibitem [{\citenamefont {Colombe}\ \emph {et~al.}(2007)\citenamefont
		{Colombe}, \citenamefont {Steinmetz}, \citenamefont {Dubois}, \citenamefont
		{Linke}, \citenamefont {Hunger},\ and\ \citenamefont
		{Reichel}}]{Colombe2007}%
	\BibitemOpen
	\bibfield  {author} {\bibinfo {author} {\bibfnamefont {Y.}~\bibnamefont
			{Colombe}}, \bibinfo {author} {\bibfnamefont {T.}~\bibnamefont {Steinmetz}},
		\bibinfo {author} {\bibfnamefont {G.}~\bibnamefont {Dubois}}, \bibinfo
		{author} {\bibfnamefont {F.}~\bibnamefont {Linke}}, \bibinfo {author}
		{\bibfnamefont {D.}~\bibnamefont {Hunger}}, \ and\ \bibinfo {author}
		{\bibfnamefont {J.}~\bibnamefont {Reichel}},\ }\href {\doibase
		10.1038/nature06331} {\bibfield  {journal} {\bibinfo  {journal} {Nature}\
		}\textbf {\bibinfo {volume} {450}},\ \bibinfo {pages} {272} (\bibinfo {year}
		{2007})}\BibitemShut {NoStop}%
	\bibitem [{\citenamefont {Steiner}\ \emph {et~al.}(2014)\citenamefont
		{Steiner}, \citenamefont {Meyer}, \citenamefont {Reichel},\ and\
		\citenamefont {K\"ohl}}]{MSteinerMeyer}%
	\BibitemOpen
	\bibfield  {author} {\bibinfo {author} {\bibfnamefont {M.}~\bibnamefont
			{Steiner}}, \bibinfo {author} {\bibfnamefont {H.~M.}\ \bibnamefont {Meyer}},
		\bibinfo {author} {\bibfnamefont {J.}~\bibnamefont {Reichel}}, \ and\
		\bibinfo {author} {\bibfnamefont {M.}~\bibnamefont {K\"ohl}},\ }\href
	{\doibase 10.1103/PhysRevLett.113.263003} {\bibfield  {journal} {\bibinfo
			{journal} {Phys. Rev. Lett.}\ }\textbf {\bibinfo {volume} {113}},\ \bibinfo
		{pages} {263003} (\bibinfo {year} {2014})}\BibitemShut {NoStop}%
	\bibitem [{\citenamefont {Takahashi}\ \emph {et~al.}(2017)\citenamefont
		{Takahashi}, \citenamefont {Kassa}, \citenamefont {Christoforou},\ and\
		\citenamefont {Keller}}]{PhysRevA.96.023824}%
	\BibitemOpen
	\bibfield  {author} {\bibinfo {author} {\bibfnamefont {H.}~\bibnamefont
			{Takahashi}}, \bibinfo {author} {\bibfnamefont {E.}~\bibnamefont {Kassa}},
		\bibinfo {author} {\bibfnamefont {C.}~\bibnamefont {Christoforou}}, \ and\
		\bibinfo {author} {\bibfnamefont {M.}~\bibnamefont {Keller}},\ }\href
	{\doibase 10.1103/PhysRevA.96.023824} {\bibfield  {journal} {\bibinfo
			{journal} {Phys. Rev. A}\ }\textbf {\bibinfo {volume} {96}},\ \bibinfo
		{pages} {023824} (\bibinfo {year} {2017})}\BibitemShut {NoStop}%
	\bibitem [{\citenamefont {Brandstätter}\ \emph {et~al.}(2013)\citenamefont
		{Brandstätter}, \citenamefont {McClung}, \citenamefont {Schüppert},
		\citenamefont {Casabone}, \citenamefont {Friebe}, \citenamefont {Stute},
		\citenamefont {Schmidt}, \citenamefont {Deutsch}, \citenamefont {Reichel},
		\citenamefont {Blatt},\ and\ \citenamefont
		{Northup}}]{doi:10.1063/1.4838696}%
	\BibitemOpen
	\bibfield  {author} {\bibinfo {author} {\bibfnamefont {B.}~\bibnamefont
			{Brandstätter}}, \bibinfo {author} {\bibfnamefont {A.}~\bibnamefont
			{McClung}}, \bibinfo {author} {\bibfnamefont {K.}~\bibnamefont {Schüppert}},
		\bibinfo {author} {\bibfnamefont {B.}~\bibnamefont {Casabone}}, \bibinfo
		{author} {\bibfnamefont {K.}~\bibnamefont {Friebe}}, \bibinfo {author}
		{\bibfnamefont {A.}~\bibnamefont {Stute}}, \bibinfo {author} {\bibfnamefont
			{P.~O.}\ \bibnamefont {Schmidt}}, \bibinfo {author} {\bibfnamefont
			{C.}~\bibnamefont {Deutsch}}, \bibinfo {author} {\bibfnamefont
			{J.}~\bibnamefont {Reichel}}, \bibinfo {author} {\bibfnamefont
			{R.}~\bibnamefont {Blatt}}, \ and\ \bibinfo {author} {\bibfnamefont {T.~E.}\
			\bibnamefont {Northup}},\ }\href {\doibase 10.1063/1.4838696} {\bibfield
		{journal} {\bibinfo  {journal} {Review of Scientific Instruments}\ }\textbf
		{\bibinfo {volume} {84}},\ \bibinfo {pages} {123104} (\bibinfo {year}
		{2013})},\ \Eprint {http://arxiv.org/abs/https://doi.org/10.1063/1.4838696}
	{https://doi.org/10.1063/1.4838696} \BibitemShut {NoStop}%
	\bibitem [{\citenamefont {Takahashi}\ \emph {et~al.}(2020)\citenamefont
		{Takahashi}, \citenamefont {Kassa}, \citenamefont {Christoforou},\ and\
		\citenamefont {Keller}}]{Takahashi2020}%
	\BibitemOpen
	\bibfield  {author} {\bibinfo {author} {\bibfnamefont {H.}~\bibnamefont
			{Takahashi}}, \bibinfo {author} {\bibfnamefont {E.}~\bibnamefont {Kassa}},
		\bibinfo {author} {\bibfnamefont {C.}~\bibnamefont {Christoforou}}, \ and\
		\bibinfo {author} {\bibfnamefont {M.}~\bibnamefont {Keller}},\ }\href
	{\doibase 10.1103/PhysRevLett.124.013602} {\bibfield  {journal} {\bibinfo
			{journal} {Phys. Rev. Lett.}\ }\textbf {\bibinfo {volume} {124}},\ \bibinfo
		{pages} {013602} (\bibinfo {year} {2020})}\BibitemShut {NoStop}%
	\bibitem [{\citenamefont {Albrecht}\ \emph {et~al.}(2013)\citenamefont
		{Albrecht}, \citenamefont {Bommer}, \citenamefont {Deutsch}, \citenamefont
		{Reichel},\ and\ \citenamefont {Becher}}]{PhysRevLett.110.243602}%
	\BibitemOpen
	\bibfield  {author} {\bibinfo {author} {\bibfnamefont {R.}~\bibnamefont
			{Albrecht}}, \bibinfo {author} {\bibfnamefont {A.}~\bibnamefont {Bommer}},
		\bibinfo {author} {\bibfnamefont {C.}~\bibnamefont {Deutsch}}, \bibinfo
		{author} {\bibfnamefont {J.}~\bibnamefont {Reichel}}, \ and\ \bibinfo
		{author} {\bibfnamefont {C.}~\bibnamefont {Becher}},\ }\href {\doibase
		10.1103/PhysRevLett.110.243602} {\bibfield  {journal} {\bibinfo  {journal}
			{Phys. Rev. Lett.}\ }\textbf {\bibinfo {volume} {110}},\ \bibinfo {pages}
		{243602} (\bibinfo {year} {2013})}\BibitemShut {NoStop}%
	\bibitem [{\citenamefont {Muller}\ \emph {et~al.}(2009)\citenamefont {Muller},
		\citenamefont {Flagg}, \citenamefont {Metcalfe}, \citenamefont {Lawall},\
		and\ \citenamefont {Solomon}}]{doi:10.1063/1.3245311}%
	\BibitemOpen
	\bibfield  {author} {\bibinfo {author} {\bibfnamefont {A.}~\bibnamefont
			{Muller}}, \bibinfo {author} {\bibfnamefont {E.~B.}\ \bibnamefont {Flagg}},
		\bibinfo {author} {\bibfnamefont {M.}~\bibnamefont {Metcalfe}}, \bibinfo
		{author} {\bibfnamefont {J.}~\bibnamefont {Lawall}}, \ and\ \bibinfo {author}
		{\bibfnamefont {G.~S.}\ \bibnamefont {Solomon}},\ }\href {\doibase
		10.1063/1.3245311} {\bibfield  {journal} {\bibinfo  {journal} {Applied
				Physics Letters}\ }\textbf {\bibinfo {volume} {95}},\ \bibinfo {pages}
		{173101} (\bibinfo {year} {2009})},\ \Eprint
	{http://arxiv.org/abs/https://doi.org/10.1063/1.3245311}
	{https://doi.org/10.1063/1.3245311} \BibitemShut {NoStop}%
	\bibitem [{\citenamefont {Ballance}\ \emph {et~al.}(2017)\citenamefont
		{Ballance}, \citenamefont {Meyer}, \citenamefont {Kobel}, \citenamefont
		{Ott}, \citenamefont {Reichel},\ and\ \citenamefont
		{K\"ohl}}]{CavityBackAction}%
	\BibitemOpen
	\bibfield  {author} {\bibinfo {author} {\bibfnamefont {T.~G.}\ \bibnamefont
			{Ballance}}, \bibinfo {author} {\bibfnamefont {H.~M.}\ \bibnamefont {Meyer}},
		\bibinfo {author} {\bibfnamefont {P.}~\bibnamefont {Kobel}}, \bibinfo
		{author} {\bibfnamefont {K.}~\bibnamefont {Ott}}, \bibinfo {author}
		{\bibfnamefont {J.}~\bibnamefont {Reichel}}, \ and\ \bibinfo {author}
		{\bibfnamefont {M.}~\bibnamefont {K\"ohl}},\ }\href {\doibase
		10.1103/PhysRevA.95.033812} {\bibfield  {journal} {\bibinfo  {journal} {Phys.
				Rev. A}\ }\textbf {\bibinfo {volume} {95}},\ \bibinfo {pages} {033812}
		(\bibinfo {year} {2017})}\BibitemShut {NoStop}%
	\bibitem [{\citenamefont {James}\ \emph {et~al.}(2001)\citenamefont {James},
		\citenamefont {Kwiat}, \citenamefont {Munro},\ and\ \citenamefont
		{White}}]{PhysRevA.64.052312}%
	\BibitemOpen
	\bibfield  {author} {\bibinfo {author} {\bibfnamefont {D.~F.~V.}\
			\bibnamefont {James}}, \bibinfo {author} {\bibfnamefont {P.~G.}\ \bibnamefont
			{Kwiat}}, \bibinfo {author} {\bibfnamefont {W.~J.}\ \bibnamefont {Munro}}, \
		and\ \bibinfo {author} {\bibfnamefont {A.~G.}\ \bibnamefont {White}},\ }\href
	{\doibase 10.1103/PhysRevA.64.052312} {\bibfield  {journal} {\bibinfo
			{journal} {Phys. Rev. A}\ }\textbf {\bibinfo {volume} {64}},\ \bibinfo
		{pages} {052312} (\bibinfo {year} {2001})}\BibitemShut {NoStop}%
	\bibitem [{\citenamefont {Bussi{\`e}res}\ \emph {et~al.}(2014)\citenamefont
		{Bussi{\`e}res}, \citenamefont {Clausen}, \citenamefont {Tiranov},
		\citenamefont {Korzh}, \citenamefont {Verma}, \citenamefont {Nam},
		\citenamefont {Marsili}, \citenamefont {Ferrier}, \citenamefont {Goldner},
		\citenamefont {Herrmann}, \citenamefont {Silberhorn}, \citenamefont {Sohler},
		\citenamefont {Afzelius},\ and\ \citenamefont {Gisin}}]{Bussieres2014}%
	\BibitemOpen
	\bibfield  {author} {\bibinfo {author} {\bibfnamefont {F.}~\bibnamefont
			{Bussi{\`e}res}}, \bibinfo {author} {\bibfnamefont {C.}~\bibnamefont
			{Clausen}}, \bibinfo {author} {\bibfnamefont {A.}~\bibnamefont {Tiranov}},
		\bibinfo {author} {\bibfnamefont {B.}~\bibnamefont {Korzh}}, \bibinfo
		{author} {\bibfnamefont {V.~B.}\ \bibnamefont {Verma}}, \bibinfo {author}
		{\bibfnamefont {S.~W.}\ \bibnamefont {Nam}}, \bibinfo {author} {\bibfnamefont
			{F.}~\bibnamefont {Marsili}}, \bibinfo {author} {\bibfnamefont
			{A.}~\bibnamefont {Ferrier}}, \bibinfo {author} {\bibfnamefont
			{P.}~\bibnamefont {Goldner}}, \bibinfo {author} {\bibfnamefont
			{H.}~\bibnamefont {Herrmann}}, \bibinfo {author} {\bibfnamefont
			{C.}~\bibnamefont {Silberhorn}}, \bibinfo {author} {\bibfnamefont
			{W.}~\bibnamefont {Sohler}}, \bibinfo {author} {\bibfnamefont
			{M.}~\bibnamefont {Afzelius}}, \ and\ \bibinfo {author} {\bibfnamefont
			{N.}~\bibnamefont {Gisin}},\ }\href {\doibase 10.1038/nphoton.2014.215}
	{\bibfield  {journal} {\bibinfo  {journal} {Nature Photonics}\ }\textbf
		{\bibinfo {volume} {8}},\ \bibinfo {pages} {775} (\bibinfo {year}
		{2014})}\BibitemShut {NoStop}%
	\bibitem [{\citenamefont {Meyer}\ \emph {et~al.}(2015)\citenamefont {Meyer},
		\citenamefont {Stockill}, \citenamefont {Steiner}, \citenamefont {Le~Gall},
		\citenamefont {Matthiesen}, \citenamefont {Clarke}, \citenamefont {Ludwig},
		\citenamefont {Reichel}, \citenamefont {Atat\"ure},\ and\ \citenamefont
		{K\"ohl}}]{IonSemiConductorSteinerMeyer}%
	\BibitemOpen
	\bibfield  {author} {\bibinfo {author} {\bibfnamefont {H.~M.}\ \bibnamefont
			{Meyer}}, \bibinfo {author} {\bibfnamefont {R.}~\bibnamefont {Stockill}},
		\bibinfo {author} {\bibfnamefont {M.}~\bibnamefont {Steiner}}, \bibinfo
		{author} {\bibfnamefont {C.}~\bibnamefont {Le~Gall}}, \bibinfo {author}
		{\bibfnamefont {C.}~\bibnamefont {Matthiesen}}, \bibinfo {author}
		{\bibfnamefont {E.}~\bibnamefont {Clarke}}, \bibinfo {author} {\bibfnamefont
			{A.}~\bibnamefont {Ludwig}}, \bibinfo {author} {\bibfnamefont
			{J.}~\bibnamefont {Reichel}}, \bibinfo {author} {\bibfnamefont
			{M.}~\bibnamefont {Atat\"ure}}, \ and\ \bibinfo {author} {\bibfnamefont
			{M.}~\bibnamefont {K\"ohl}},\ }\href@noop {} {\bibfield  {journal} {\bibinfo
			{journal} {Phys. Rev. Lett.}\ }\textbf {\bibinfo {volume} {114}},\ \bibinfo
		{pages} {123001} (\bibinfo {year} {2015})}\BibitemShut {NoStop}%
	\bibitem [{\citenamefont {Olmschenk}\ \emph {et~al.}(2007)\citenamefont
		{Olmschenk}, \citenamefont {Younge}, \citenamefont {Moehring}, \citenamefont
		{Matsukevich}, \citenamefont {Maunz},\ and\ \citenamefont
		{Monroe}}]{PhysRevA.76.052314}%
	\BibitemOpen
	\bibfield  {author} {\bibinfo {author} {\bibfnamefont {S.}~\bibnamefont
			{Olmschenk}}, \bibinfo {author} {\bibfnamefont {K.~C.}\ \bibnamefont
			{Younge}}, \bibinfo {author} {\bibfnamefont {D.~L.}\ \bibnamefont
			{Moehring}}, \bibinfo {author} {\bibfnamefont {D.~N.}\ \bibnamefont
			{Matsukevich}}, \bibinfo {author} {\bibfnamefont {P.}~\bibnamefont {Maunz}},
		\ and\ \bibinfo {author} {\bibfnamefont {C.}~\bibnamefont {Monroe}},\ }\href
	{\doibase 10.1103/PhysRevA.76.052314} {\bibfield  {journal} {\bibinfo
			{journal} {Phys. Rev. A}\ }\textbf {\bibinfo {volume} {76}},\ \bibinfo
		{pages} {052314} (\bibinfo {year} {2007})}\BibitemShut {NoStop}%
	\bibitem [{\citenamefont {Ikuta}\ \emph {et~al.}(2018)\citenamefont {Ikuta},
		\citenamefont {Kobayashi}, \citenamefont {Kawakami}, \citenamefont {Miki},
		\citenamefont {Yabuno}, \citenamefont {Yamashita}, \citenamefont {Terai},
		\citenamefont {Koashi}, \citenamefont {Mukai}, \citenamefont {Yamamoto},\
		and\ \citenamefont {Imoto}}]{Ikuta2018}%
	\BibitemOpen
	\bibfield  {author} {\bibinfo {author} {\bibfnamefont {R.}~\bibnamefont
			{Ikuta}}, \bibinfo {author} {\bibfnamefont {T.}~\bibnamefont {Kobayashi}},
		\bibinfo {author} {\bibfnamefont {T.}~\bibnamefont {Kawakami}}, \bibinfo
		{author} {\bibfnamefont {S.}~\bibnamefont {Miki}}, \bibinfo {author}
		{\bibfnamefont {M.}~\bibnamefont {Yabuno}}, \bibinfo {author} {\bibfnamefont
			{T.}~\bibnamefont {Yamashita}}, \bibinfo {author} {\bibfnamefont
			{H.}~\bibnamefont {Terai}}, \bibinfo {author} {\bibfnamefont
			{M.}~\bibnamefont {Koashi}}, \bibinfo {author} {\bibfnamefont
			{T.}~\bibnamefont {Mukai}}, \bibinfo {author} {\bibfnamefont
			{T.}~\bibnamefont {Yamamoto}}, \ and\ \bibinfo {author} {\bibfnamefont
			{N.}~\bibnamefont {Imoto}},\ }\href {\doibase 10.1038/s41467-018-04338-x}
	{\bibfield  {journal} {\bibinfo  {journal} {Nature Communications}\ }\textbf
		{\bibinfo {volume} {9}},\ \bibinfo {pages} {1997} (\bibinfo {year}
		{2018})}\BibitemShut {NoStop}%
	\bibitem [{\citenamefont {R\"utz}\ \emph {et~al.}(2017)\citenamefont {R\"utz},
		\citenamefont {Luo}, \citenamefont {Suche},\ and\ \citenamefont
		{Silberhorn}}]{PhysRevApplied.7.024021}%
	\BibitemOpen
	\bibfield  {author} {\bibinfo {author} {\bibfnamefont {H.}~\bibnamefont
			{R\"utz}}, \bibinfo {author} {\bibfnamefont {K.-H.}\ \bibnamefont {Luo}},
		\bibinfo {author} {\bibfnamefont {H.}~\bibnamefont {Suche}}, \ and\ \bibinfo
		{author} {\bibfnamefont {C.}~\bibnamefont {Silberhorn}},\ }\href {\doibase
		10.1103/PhysRevApplied.7.024021} {\bibfield  {journal} {\bibinfo  {journal}
			{Phys. Rev. Applied}\ }\textbf {\bibinfo {volume} {7}},\ \bibinfo {pages}
		{024021} (\bibinfo {year} {2017})}\BibitemShut {NoStop}%
	\bibitem [{\citenamefont {Kasture}\ \emph {et~al.}(2016)\citenamefont
		{Kasture}, \citenamefont {Lenzini}, \citenamefont {Haylock}, \citenamefont
		{Boes}, \citenamefont {Mitchell}, \citenamefont {Streed},\ and\ \citenamefont
		{Lobino}}]{Kasture_2016}%
	\BibitemOpen
	\bibfield  {author} {\bibinfo {author} {\bibfnamefont {S.}~\bibnamefont
			{Kasture}}, \bibinfo {author} {\bibfnamefont {F.}~\bibnamefont {Lenzini}},
		\bibinfo {author} {\bibfnamefont {B.}~\bibnamefont {Haylock}}, \bibinfo
		{author} {\bibfnamefont {A.}~\bibnamefont {Boes}}, \bibinfo {author}
		{\bibfnamefont {A.}~\bibnamefont {Mitchell}}, \bibinfo {author}
		{\bibfnamefont {E.~W.}\ \bibnamefont {Streed}}, \ and\ \bibinfo {author}
		{\bibfnamefont {M.}~\bibnamefont {Lobino}},\ }\href {\doibase
		10.1088/2040-8978/18/10/104007} {\bibfield  {journal} {\bibinfo  {journal}
			{Journal of Optics}\ }\textbf {\bibinfo {volume} {18}},\ \bibinfo {pages}
		{104007} (\bibinfo {year} {2016})}\BibitemShut {NoStop}%
	\bibitem [{\citenamefont {Schmitz}\ \emph {et~al.}(2019)\citenamefont
		{Schmitz}, \citenamefont {Meyer},\ and\ \citenamefont
		{Köhl}}]{doi:10.1063/1.5093551}%
	\BibitemOpen
	\bibfield  {author} {\bibinfo {author} {\bibfnamefont {J.}~\bibnamefont
			{Schmitz}}, \bibinfo {author} {\bibfnamefont {H.~M.}\ \bibnamefont {Meyer}},
		\ and\ \bibinfo {author} {\bibfnamefont {M.}~\bibnamefont {Köhl}},\ }\href
	{\doibase 10.1063/1.5093551} {\bibfield  {journal} {\bibinfo  {journal}
			{Review of Scientific Instruments}\ }\textbf {\bibinfo {volume} {90}},\
		\bibinfo {pages} {063102} (\bibinfo {year} {2019})},\ \Eprint
	{http://arxiv.org/abs/https://doi.org/10.1063/1.5093551}
	{https://doi.org/10.1063/1.5093551} \BibitemShut {NoStop}%
	\bibitem [{\citenamefont {Harlander}\ \emph {et~al.}(2010)\citenamefont
		{Harlander}, \citenamefont {Brownnutt}, \citenamefont {Hänsel},\ and\
		\citenamefont {Blatt}}]{DielectricChargingUV}%
	\BibitemOpen
	\bibfield  {author} {\bibinfo {author} {\bibfnamefont {M.}~\bibnamefont
			{Harlander}}, \bibinfo {author} {\bibfnamefont {M.}~\bibnamefont
			{Brownnutt}}, \bibinfo {author} {\bibfnamefont {W.}~\bibnamefont {Hänsel}},
		\ and\ \bibinfo {author} {\bibfnamefont {R.}~\bibnamefont {Blatt}},\ }\href
	{\doibase 10.1088/1367-2630/12/9/093035} {\bibfield  {journal} {\bibinfo
			{journal} {New Journal of Physics}\ }\textbf {\bibinfo {volume} {12}},\
		\bibinfo {pages} {093035} (\bibinfo {year} {2010})}\BibitemShut {NoStop}%
	\bibitem [{\citenamefont {Vasconcelos}\ \emph {et~al.}(2020)\citenamefont
		{Vasconcelos}, \citenamefont {Reisenbauer}, \citenamefont {Salter},
		\citenamefont {Wachter}, \citenamefont {Wirtitsch}, \citenamefont
		{Schmiedmayer}, \citenamefont {Walther},\ and\ \citenamefont
		{Trupke}}]{Vasconcelos2020}%
	\BibitemOpen
	\bibfield  {author} {\bibinfo {author} {\bibfnamefont {R.}~\bibnamefont
			{Vasconcelos}}, \bibinfo {author} {\bibfnamefont {S.}~\bibnamefont
			{Reisenbauer}}, \bibinfo {author} {\bibfnamefont {C.}~\bibnamefont {Salter}},
		\bibinfo {author} {\bibfnamefont {G.}~\bibnamefont {Wachter}}, \bibinfo
		{author} {\bibfnamefont {D.}~\bibnamefont {Wirtitsch}}, \bibinfo {author}
		{\bibfnamefont {J.}~\bibnamefont {Schmiedmayer}}, \bibinfo {author}
		{\bibfnamefont {P.}~\bibnamefont {Walther}}, \ and\ \bibinfo {author}
		{\bibfnamefont {M.}~\bibnamefont {Trupke}},\ }\href {\doibase
		10.1038/s41534-019-0236-x} {\bibfield  {journal} {\bibinfo  {journal} {npj
				Quantum Information}\ }\textbf {\bibinfo {volume} {6}},\ \bibinfo {pages} {9}
		(\bibinfo {year} {2020})}\BibitemShut {NoStop}%
	\bibitem [{Note1()}]{Note1}%
	\BibitemOpen
	\bibinfo {note} {MCL61}\BibitemShut {NoStop}%
	\bibitem [{\citenamefont {Berkeland}\ and\ \citenamefont
		{Boshier}(2002)}]{PhysRevA.65.033413}%
	\BibitemOpen
	\bibfield  {author} {\bibinfo {author} {\bibfnamefont {D.~J.}\ \bibnamefont
			{Berkeland}}\ and\ \bibinfo {author} {\bibfnamefont {M.~G.}\ \bibnamefont
			{Boshier}},\ }\href {\doibase 10.1103/PhysRevA.65.033413} {\bibfield
		{journal} {\bibinfo  {journal} {Phys. Rev. A}\ }\textbf {\bibinfo {volume}
			{65}},\ \bibinfo {pages} {033413} (\bibinfo {year} {2002})}\BibitemShut
	{NoStop}%
	\bibitem [{\citenamefont {Milner}\ and\ \citenamefont
		{Prior}(1998)}]{PhysRevLett.80.940}%
	\BibitemOpen
	\bibfield  {author} {\bibinfo {author} {\bibfnamefont {V.}~\bibnamefont
			{Milner}}\ and\ \bibinfo {author} {\bibfnamefont {Y.}~\bibnamefont {Prior}},\
	}\href {\doibase 10.1103/PhysRevLett.80.940} {\bibfield  {journal} {\bibinfo
			{journal} {Phys. Rev. Lett.}\ }\textbf {\bibinfo {volume} {80}},\ \bibinfo
		{pages} {940} (\bibinfo {year} {1998})}\BibitemShut {NoStop}%
\end{thebibliography}
\end{document}